\documentclass[12pt,a4paper]{article}
\usepackage{amsmath,amsfonts,amsthm,graphicx,lscape}
\usepackage[dvips]{psfrag}
\usepackage{cite}

\usepackage{textcomp}
\usepackage{amsmath,amsthm,amssymb}
\usepackage{amsfonts,graphicx,lscape}

\numberwithin{equation}{section}

\theoremstyle{definition}
\newtheorem{theorem}{Theorem}[section]

\newtheorem{proposition}[theorem]{Proposition}

\newcommand{\vt}[1]{\mbox{\boldmath$#1$}}

\begin{document}
\title{
\vspace{-9mm}
Systematic method of
generating 
new integrable systems 
via 
inverse Miura maps 
}
\author{Takayuki \textsc{Tsuchida}\footnote{
E-mail:\ surname at 
ms.u-tokyo.ac.jp
}
\\
\\
{\it Okayama Institute for Quantum Physics,
}
\\
{\it Kyoyama 1-9-1, Okayama 700-0015, Japan}
}
\maketitle
\begin{abstract} 
We 
provide a new 
natural 
interpretation of the 
%
Lax representation for an integrable system; 
that is, 
the spectral problem is 
the linearized form of a Miura transformation between 
the original system 
and 
a modified version of it. 
On the basis of 
this interpretation, we formulate a 
systematic method 
of identifying
modified integrable systems 
that can be mapped to a given 
integrable system 
by Miura transformations.
Thus, 
this method can be used to 
generate new integrable systems 
from known 
systems 
through
inverse Miura maps; 
it can be applied to 
both continuous and discrete 
systems in \mbox{$1+1$} 
dimensions
as well as 
in 
\mbox{$2+1$} dimensions. 
The effectiveness 
of 
the 
method 
is illustrated 
using examples such as 
the nonlinear Schr\"odinger (NLS) system, 
the Zakharov--Ito 
system (two-component KdV), 
the three-wave interaction system, 
the Yajima--Oikawa system, 
the Ablowitz--Ladik lattice 
(integrable space-discrete NLS),
and two \mbox{$(2+1)$}-dimensional NLS systems. 
\end{abstract}
%
\vspace{5mm}
\mbox{$(2+1)$}-dimensional NLS,
integrable lattices, 
derivative three-wave interaction system, 
derivative Yajima--Oikawa system
\vspace{5mm}
%
%
%
\newpage
\noindent
\tableofcontents

\newpage
\section{Introduction}

For an 
integrable system that is essentially nonlinear 
and thus 
is 
not linearizable by a change of variables, 
the Lax (or zero-curvature) representation~\cite{Lax} 
occupies a central position 
in 
its integrability properties; 
giving 
the Lax representation is even 
considered as 
proof of integrability. 
The first example of a Lax representation was found
in the late 
60s for the Korteweg--de Vries (KdV) 
equation~\cite{GGKM}; it was originally derived 
through 
linearizing 
a Riccati-type 
transformation from 
the modified KdV (mKdV) equation 
(or, more generally, 
its one-parameter 
generalization called the Gardner equation~\cite{Miura76}) 
to the KdV equation~\cite{Miura68,MGK68}. 
Such a nontrivial 
and non-ultralocal 
transformation of dependent variables 
from one 
integrable system to another 
is now called a Miura map~\cite{Miura76,Miura68}. 
Subsequently, 
in the early 70s, the Lax representation 
in \mbox{$2 \times 2$} matrix form  
was found
for the nonlinear Schr\"odinger (NLS) equation~\cite{ZS1}, 
without any reference to 
the relevant Miura map. 
Since then, 
the number of integrable systems admitting the Lax representation 
has 
been increasing 
rapidly, 
including 
multicomponent systems, 
higher dimensional systems~\cite{AC91,Kono92,Kono93}, 
and discrete systems~\cite{Suris03}. 
However, as the Lax representations have been extended 
in various directions, the primary 
role played by the original Miura map in the KdV case 
has gotten lost in oblivion. 

The main theme of this paper is to demonstrate,
using 
an abundance of 
specific examples, that the Miura maps are 
by no means less important than the 
Lax representations.
In fact, a Miura map can generally produce
the relevant Lax 
representation and vice versa. 
Thus, they are different facets of the same 
property 
and, 
in a 
sense, equivalent. 
This is not merely 
a 
conceptual integration of 
the Miura maps and the Lax representations; 
rather, we 
will 
use it for a more practical and attractive purpose, 
that is, to generate new integrable systems from known systems 
in a systematic manner. 
Actually, 
no such methods have ever 
been proposed 
and 
applied 
successfully 
to a broad 
spectrum of 
examples. 
The basic idea of our 
new method 
can be 
best described
%
using 
the nonreduced NLS system~\cite{AKNS}
%
%
\begin{subequations} 
\label{mNLS}
\begin{align}
& \mathrm{i} Q_t + Q_{xx} - 2Q R Q = O, 
\label{NLS1}
\\
& \mathrm{i} R_t - R_{xx} + 2R Q R = O,
\label{NLS2}
\end{align}
\end{subequations}
as an illustrative example. 
The subscripts $t$ and $x$ denote the partial
differentiation with respect to these variables.
Note that 
the NLS system (\ref{mNLS}) is integrable for matrix-valued 
dependent variables~\cite{ZS3}; 
in the 
general case, 
$Q$ and $R$ are 
\mbox{$l_1 \times l_2$} 
and \mbox{$l_2 \times l_1$} matrices, respectively. 
In this paper, 
the symbol
$O$
on the right-hand side
of the
equations
implies that the dependent variables can take
their
values in
matrices.
The Lax representation 
for the matrix NLS system (\ref{mNLS})
is given by~\cite{Zakh,Konop1}
\begin{subequations}
\label{NLS-UV}
\begin{align}
& \left[
\begin{array}{c}
 \Psi_1  \\
 \Psi_2 \\
\end{array}
\right]_x 
= \left[
\begin{array}{cc}
-\mathrm{i}\zeta I_1 & Q \\
 R & \mathrm{i}\zeta I_2\\
\end{array}
\right] 
\left[
\begin{array}{c}
 \Psi_1  \\
 \Psi_2 \\
\end{array}
\right],
\label{NLS-U}
\\[1mm]
& \left[
\begin{array}{c}
 \Psi_1  \\
 \Psi_2 \\
\end{array}
\right]_t 
= \left[
\begin{array}{cc}
-2\mathrm{i}\zeta^2 I_1 -\mathrm{i} QR & 2 \zeta Q + \mathrm{i} Q_x \\
 2 \zeta R - \mathrm{i} R_x & 2\mathrm{i}\zeta^2 I_2 +\mathrm{i} RQ \\
\end{array}
\right]
\left[
\begin{array}{c}
 \Psi_1  \\
 \Psi_2 \\
\end{array}
\right].
\label{NLS-V}
\end{align}
\end{subequations}
Here, 
$\zeta$ is the spectral parameter, which is
an arbitrary constant
independent of $x$ and $t$, and 
$I_1$ and $I_2$ 
are the \mbox{$l_1 \times l_1$} and \mbox{$l_2 \times l_2$} 
unit matrices, respectively. 
We consider an \mbox{$(l_1+l_2) \times l_1$} 
matrix-valued solution 
for
the pair of linear equations (\ref{NLS-UV}) 
such that $\Psi_1$ is an \mbox{$l_1 \times l_1$} 
invertible 
matrix. 
Then, 
in terms of 
the \mbox{$l_2 \times l_1$} matrix \mbox{$P := \Psi_2 \Psi_1^{-1}$}, 
(\ref{NLS-UV}) can be 
rewritten as a pair of matrix Riccati equations 
(see~\cite{Chen1,Chen2,WSK} for the scalar case), 
\begin{subequations}
\label{NLS-R}
\begin{align}
& P_x = R+ 2 \mathrm{i} \zeta P -PQP, 
\label{NLS-R1}
\\
& P_t = 2 \zeta R- \mathrm{i}R_x + 4 \mathrm{i}\zeta^2 P + \mathrm{i}RQ P 
	+ \mathrm{i}PQR - 2\zeta PQP  - \mathrm{i}PQ_x P.
\label{NLS-R2}
\end{align}
\end{subequations}
%
Using (\ref{NLS-R1}), 
we can 
express 
$R$ 
in terms of $P$ and $Q$ 
as \mbox{$ -2\mathrm{i} \zeta P +P_x +PQP$}. 
Thus, 
(\ref{NLS1}) and (\ref{NLS-R2}) 
now comprise 
a closed system for $Q$ and $P$, i.e., 
\begin{subequations}
\label{mGI}
\begin{align}
& \mathrm{i} Q_t + Q_{xx} +4\mathrm{i} \zeta QPQ - 2Q P_x Q -2QPQPQ = O,
\label{GI1}
\\
& \mathrm{i} P_t - P_{xx} -4\mathrm{i} \zeta PQP - 2P Q_x P +2PQPQP = O.
\label{GI2}
\end{align}
\end{subequations}
%
This is 
intrinsically a derivative NLS system, 
which was investigated 
by Ablowitz {\it et al.}~\cite{ARS}
and Gerdjikov and Ivanov~\cite{GI}
in the case of scalar dependent variables. 
The matrix generalization (\ref{mGI}) was studied 
in~\cite{Linden1,Olver2,TW3,Ad,Dimakis}. 
Note that the free parameter $\zeta$ 
is nonessential 
as long as we 
consider the Gerdjikov--Ivanov system 
(\ref{mGI}) 
separately 
as an isolated 
system; 
indeed, it can be set equal to zero 
using a Galilean transformation (cf.~\cite{IKWS,Kawata2}).
%
It is 
straightforward to 
confirm that 
if the pair 
\mbox{$(Q,P)$} 
satisfies (\ref{mGI}), then 
the pair 
\mbox{$(Q,R)$} 
with \mbox{$ R :=-2\mathrm{i} \zeta P +P_x +PQP$} 
indeed 
satisfies (\ref{mNLS});
this fact is already 
known in the case \mbox{$\zeta=0$} 
(see 
\cite{GI,KN,GIK,WS} for the scalar case 
and \cite{Linden1,Ad} 
for the matrix case). 
Thus, starting with the 
NLS system (\ref{mNLS}) and 
its Lax representation (\ref{NLS-UV}), 
we 
obtain a 
Miura map 
from the 
Gerdjikov--Ivanov system (\ref{mGI}) to the 
NLS system (\ref{mNLS}), 
without 
any prior knowledge
%
%
about 
(\ref{mGI}). 
The 
spectral problem (\ref{NLS-U}) 
turns out to be the linearized form of 
the 
defining 
relation (\ref{NLS-R1}) 
for
the Miura map 
(cf.~\cite{Kono82,Calo84}); 
%
%
the entire Lax representation (\ref{NLS-UV}) 
not only yields 
the NLS system (\ref{mNLS}) 
as 
the compatibility condition 
but also 
enables us
to 
apply 
an {\it inverse Miura map} \/to 
(\ref{mNLS}). 
Somewhat similar 
computations 
were given in~\cite{chowd94,Bori01}, 
but they are not as 
straightforward 
and 
comprehensible as 
ours. 
The 
``nonessential'' parameter $\zeta$ plays a crucial role 
when studying 
the original system (\ref{mNLS}) and the modified system (\ref{mGI}) 
in a unified way. 
For instance, 
the simplest 
conservation law for (\ref{mGI}), 
\begin{equation}
\mathrm{i} 
\frac{\partial}{\partial t} \mathrm{tr}\hspace{1pt}(QP)
 + \frac{\partial}{\partial x} 
	\mathrm{tr}\left[ Q_{x}P-QP_x -(QP)^2 \right]
= 0,
\label{GI-cons}
\end{equation}
works as a generating function 
of 
an infinite set of conservation laws for (\ref{mNLS}). 
Indeed, 
following 
the same 
routine 
as 
in the 
case of the 
original Miura map~\cite{Miura76,MGK68}, 
we can formally express $P$ 
using (\ref{NLS-R1}) 
in 
a power series of $(2 \mathrm{i}\zeta)^{-1}$, 
\begin{equation}
P= -\frac{R}{2 \mathrm{i}\zeta} 
 -\frac{R_x }{(2 \mathrm{i}\zeta)^2} 
 -\frac{R_{xx} -RQR}{(2 \mathrm{i}\zeta)^3} 
+ \cdots. 
\label{P-series}
\end{equation}
Thus, 
substituting (\ref{P-series}) into (\ref{GI-cons}) and 
equating 
all the coefficients of different powers of 
$(2 \mathrm{i}\zeta)^{-1}$ with zero, we obtain 
an infinite set of conservation laws for the NLS system (\ref{mNLS});
this uncovers 
%
the hidden essential 
meaning
of the method 
devised 
by Wadati
{\it et al.}~\cite{WSK} (also see~\cite{Haberman}) 
for
constructing 
conservation laws 
recursively. 
Then, 
using (\ref{NLS-R1}), namely, 
\mbox{$ R =-2\mathrm{i} \zeta P +P_x +PQP$},
we can 
also 
obtain 
an infinite set of conservation laws for the 
Gerdjikov--Ivanov system (\ref{mGI}). 
%
Moreover, using the inverse scattering method based on 
the Lax representation (\ref{NLS-UV}), 
it is possible to derive the solution formulas for 
the NLS system (\ref{mNLS}) and 
for the Gerdjikov--Ivanov system (\ref{mGI}) 
concurrently. 
%

Our method described 
above actually applies to a surprisingly 
wide spectrum of integrable systems and is not limited
to 
systems of two coupled partial differential equations (PDEs) 
in \mbox{$1+1$} space-time 
dimensions, such as (\ref{mNLS}). 
Broadly speaking, 
the
method is applicable 
to every
integrable system with a Lax representation, 
the spatial part of which 
is {\it ultralocal} \/in the 
dependent variables; 
then, 
the method successfully 
identifies modified integrable systems 
embedded in the original system via Miura maps. 
Thus, it can be applied to most of the known integrable systems, 
including 
multicomponent PDEs,
\mbox{$(2+1)$}-dimensional PDEs, 
differential-difference equations, 
and partial difference equations. 
%
%
%
%
The Lax representation 
for an integrable system 
can be determined only up to 
gauge transformations; 
we must
fix the gauge 
appropriately 
to write the spectral problem 
in an ultralocal form. 
For instance, the Gerdjikov--Ivanov system (\ref{mGI}) 
with \mbox{$\zeta=\zeta_1$} 
allows
the nonstandard 
Lax representation, namely, 
(\ref{NLS-UV}) 
with \mbox{$ R =-2\mathrm{i} \zeta_1 P +P_x +PQP$}; 
this is not a convenient form 
for applying our method, 
because the spectral problem is not ultralocal in $P$. 
Thus, before 
applying the
method, 
we 
consider 
a suitable gauge transformation, 
{\it e.g.}, \mbox{$\Phi_1:= \Psi_1$} and \mbox{$\Phi_2:= \Psi_2 -P \Psi_1$}, 
so that the spectral problem assumes
an ultralocal form with respect to 
both $Q$ and $P$.
However, such a procedure is 
not always possible 
for a 
nonstandard spectral problem 
and our method can, in general, 
be applied 
to a given integrable system 
with only a finite number of iterations.
That is, it terminates at some stage, 
resulting in 
a finite 
chain of 
$j$-th modified
systems 
(\mbox{$j=0,1,
\ldots, j_{\mathrm{max}}$}). 

The remainder of this paper is organized as follows. 
In section 2, 
we describe 
the 
method in a general setting. 
First, we reinterpret the spatial part of a
given Lax representation 
as the defining relations of 
a Miura map;
subsequently, the temporal part of the Lax representation 
is used to 
emboss
an a priori unknown
modified 
system related to the original system by the Miura map. 
For an integrable system in \mbox{$1+1$} 
dimensions, 
the spatial part of the Lax representation is 
genuinely a spectral problem, 
that is, 
it 
involves intrinsically
the 
spectral parameter, 
which is an arbitrary constant. 
Thus, the relevant Miura map and 
modified 
system 
always contain 
an additional
free parameter
that the original system does not have. 
Using this parameter as an expansion parameter, 
the simplest conservation laws for the modified 
system 
can generate 
infinitely many conservation laws for the original system,  
which in turn 
provide infinitely many conservation laws 
for the modified system. 
Sections 3 and 4 
are devoted to 
specific examples of 
such \mbox{$(1+1)$}-dimensional integrable systems. 
In section 3, we apply the method to 
continuous systems, namely, PDEs 
such as 
the Zakharov--Ito system~\cite{Zakh,Ito},  
which 
is a two-component extension of the KdV equation, 
the Jaulent--Miodek system~\cite{JM}, 
the three-wave interaction system~\cite{ZM73},
and the Yajima--Oikawa system~\cite{YO}. 
In section 4, we 
investigate space-discrete systems; 
the Toda lattice in Flaschka--Manakov 
coordinates~\cite{Flaschka1,Flaschka2,Manakov74}, 
the Belov--Chaltikian lattice~\cite{BC1}, 
the relativistic Toda lattice~\cite{Rui90}, 
and the Ablowitz--Ladik lattice~\cite{AL1} 
are discussed 
as illustrative examples. 
In section 5, we consider 
integrable 
PDEs in \mbox{$2+1$} 
dimensions. 
Their Lax representations 
can be 
divided 
into two 
distinct 
groups 
according to the dimensionality 
of 
the spatial 
Lax operators~\cite{Zakh}. 
In the first group, the spatial 
part of a Lax representation 
is a one-dimensional spectral problem 
as in the \mbox{$(1+1)$}-dimensional case~\cite{Calo75,Calo76,Calo82};
only 
the associated 
time evolution 
exhibits 
a 
\mbox{$(2+1)$}-dimensional 
nature. 
In the second group, the spatial 
part of a Lax representation 
is genuinely a two-dimensional problem; 
in contrast to the \mbox{$(1+1)$}-dimensional case, 
it does not contain any essential spectral parameter. 
Our method can be applied to both groups, 
but in 
the latter case, 
the resulting Miura map and modified integrable system 
contain 
no additional 
parameter. 
We take
one representative example 
from each group, 
namely, the Calogero--Degasperis system~\cite{Calo76} 
and the Davey--Stewartson system~\cite{DS74}, 
to illustrate the method; 
they are both 
\mbox{$(2+1)$}-dimensional generalizations 
of the NLS system (\ref{mNLS}). 
Actually, 
our method is also applicable to discrete systems in 
\mbox{$2+1$} dimensions. However, the examples 
that we know 
appear to be too complicated to 
discuss here. 
Section 6 is devoted to concluding remarks. 
In the appendices, 
we present a natural spin-off version of our method 
using 
illustrative examples. 
In appendix~\ref{app1}, 
we prove that the new pair of dependent variables 
\mbox{$(\Psi_1^{-1} Q, \hspace{1pt}\Psi_2)$}
defined from 
the Lax representation (\ref{NLS-UV}) for 
the 
NLS system (\ref{mNLS})
satisfies 
a closed system, which is equivalent to a
derivative NLS system, 
called 
the Chen--Lee--Liu system~\cite{CLL}. 
In appendix~\ref{app2} and appendix~\ref{app3}, 
we obtain analogous results 
for the discrete case and 
the \mbox{$(2+1)$}-dimensional case, 
respectively. 
%

%
%
%

%
%

\section{Method: from 
Lax representation 
to Miura map}
\label{sect2}

In this section, we 
describe the 
essence of our method
using a general 
Lax representation
in which the
spatial part is 
a one-dimensional matrix spectral problem that 
is first order in 
the 
space 
variable. 
It can be easily modified/extended 
to the case where 
the spatial part of a Lax representation is 
a higher-order scalar problem or 
a two-dimensional matrix problem; 
some specific examples will be 
considered 
in subsequent sections. 

We consider a pair of linear equations for a column-vector function
$\vt{\psi}$,
\begin{equation}
\widehat{T}_1 \vt{\psi} = U(\zeta) \vt{\psi},
	\hspace{5mm} \widehat{T}_2 \vt{\psi} = V(\zeta)\vt{\psi}.
\label{lin1}
\end{equation}
Here, \mbox{$\widehat{T}_j \; (j=1,2)$}
are linear operators and 
\mbox{$U(\zeta)$} and \mbox{$V(\zeta)$} are square matrices
depending on the spectral parameter $\zeta$. 
Note that 
in the \mbox{$(2+1)$}-dimensional case, 
$\zeta$ is not required to be a constant; 
it 
is allowed to depend on 
some independent variables, 
as 
illustrated 
in subsection~\ref{sec-Zak}. 
The first equation in (\ref{lin1}) is 
the spatial part of the Lax representation, wherein
the linear operator $\widehat{T}_1$
denotes 
the partial 
differentiation by $x$ 
\mbox{$(\hspace{1pt}\widehat{T}_1 \vt{\psi} 
	:= \partial_x \vt{\psi}\hspace{1pt})$}
in the continuous-space case 
or the forward shift operator 
\mbox{$(\hspace{1pt}\widehat{T}_1 \vt{\psi}_n 
	:= \vt{\psi}_{n+1}\hspace{1pt})$}
in the discrete-space case. 
Similarly, the second equation in (\ref{lin1}) is 
the temporal part of the Lax representation;
in the \mbox{$(1+1)$}-dimensional case, 
$\widehat{T}_2$ denotes 
the partial 
differentiation by $t$ 
\mbox{$(\hspace{1pt}\widehat{T}_2 \vt{\psi} 
	:= \partial_t \vt{\psi}\hspace{1pt})$}
in the continuous-time case 
or the forward shift operator 
\mbox{$(\hspace{1pt}\widehat{T}_2 \vt{\psi}_{n,m} 
	:= \vt{\psi}_{n,m+1}
\hspace{1pt})$}
in the discrete-time case. 
In the \mbox{$(2+1)$}-dimensional case, 
$\widehat{T}_2$ denotes a suitable linear combination
of partial differential operators, {\it e.g.}, 
\mbox{$\widehat{T}_2 = \partial_t + f(\zeta) 
\partial_y$}~\cite{Bogo91,Str92,Stra92}. 
It is more convenient to write (\ref{lin1}) 
explicitly 
in 
the component forms 
\begin{subequations}
\label{lin2}
\begin{align}
\widehat{T}_1 
\left[
\begin{array}{c}
\Psi_1 \\
\vdots \\
\Psi_l \\
\end{array}
\right] & = \left[
\begin{array}{ccc}
U_{11} & \cdots & U_{1l}\\
\vdots & \ddots & \vdots \\
U_{l1} & \cdots & U_{l\hspace{1pt}l} \\
\end{array}
\right]
\left[
\begin{array}{c}
\Psi_1 \\
\vdots \\
\Psi_l \\
\end{array}
\right],
\label{lin2-1}
\\[3mm]
\widehat{T}_2 
\left[
\begin{array}{c}
\Psi_1 \\
\vdots \\
\Psi_l \\
\end{array}
\right] & = \left[
\begin{array}{ccc}
V_{11} & \cdots & V_{1l}\\
\vdots & \ddots & \vdots \\
V_{l1} & \cdots & V_{l\hspace{1pt}l} \\
\end{array}
\right]
\left[
\begin{array}{c}
\Psi_1 \\
\vdots \\
\Psi_l \\
\end{array}
\right],
\label{lin2-2}
\end{align}
\end{subequations}
where \mbox{$l \ge 2$}. 
The entries of the matrices 
\mbox{$U$} and \mbox{$V$} are 
classical quantities, i.e., 
$U_{ij}$ and $V_{ij}$ do not involve operators; however, 
they can
take their values in 
submatrices if \mbox{$U$} and \mbox{$V$} are 
broken into blocks. 
In such a case, we collect and 
align a set of linearly independent column-vector solutions 
of (\ref{lin2}) so that some 
$\Psi_j$ 
become
invertible 
matrices
if needed (see below). 

The compatibility condition 
\mbox{$\widehat{T}_1 \widehat{T}_2 \vt{\psi} = 
	\widehat{T}_2 \widehat{T}_1 \vt{\psi}$}
for the overdetermined system (\ref{lin1}) 
results in 
some nontrivial 
relation 
between \mbox{$U$} and \mbox{$V$}; 
this relation is 
often referred to as the zero-curvature 
condition. 
For example, in the continuous 
\mbox{$(1+1)$}-dimensional case, 
we take
\mbox{$\widehat{T}_1 = \partial_x$} and \mbox{$\widehat{T}_2 = \partial_t$}
to obtain
the equation~\cite{AKNS74,ZS79}
\[
 U_t - V_x + 
	UV-VU =O.
\]
%
%
%
If we specify 
the
$\zeta$-dependent
matrices $U$ and $V$ 
appropriately,
the zero-curvature condition 
provides 
a closed nonlinear 
system 
for some 
$\zeta$-independent 
quantities in 
$U$ and 
$V$ (cf.\ (\ref{mNLS}) and (\ref{NLS-UV})). 
In such 
a case, 
(\ref{lin1}) 
gives 
the Lax representation for the integrable system. 
For brevity, 
we assume that the spatial Lax matrix $U$ is 
{\it ultralocal} \/in the dependent variables; 
that is, if $U$ involves 
some 
dependent 
variable, 
then 
$U$ does not 
involve its 
derivatives 
and 
shifts with respect to the independent variables. 
%
In this section, 
we express 
the entire set of functionally independent dynamical variables 
contained in $U$ as 
\mbox{$\{q_1, q_2, \ldots, q_N\}$}; 
these $N$ variables may take their values in matrices. 
Although our method is 
also applicable to 
nonevolutionary systems 
that are 
negative flows of integrable hierarchies, 
for notational convenience, 
we discuss 
the simpler case of 
evolutionary systems. 
Thus, 
the time evolution of 
the integrable system considered 
can be written explicitly as 
%
\begin{subequations}
\label{evo1}
\begin{equation}
\partial_t q_i = f_i (q_1, q_2, \ldots, q_N), 
\hspace{5mm} i=1, 2, \ldots, N
\label{}
\end{equation}
in the continuous-time case 
and 
\begin{equation}
\widetilde{q}_i = f_i (q_1, q_2, \ldots, q_N), 
\hspace{5mm} i=1, 2, \ldots, N
\label{}
\end{equation}
\end{subequations}
in the discrete-time case. 
Here, the tilde
denotes
the forward
shift (\mbox{$m \to m+1$})
in the discrete-time coordinate \mbox{$m \in {\mathbb Z}$}. 
Note that the functions $f_i$ on the right-hand side of (\ref{evo1}) 
are 
not ultralocal in their arguments 
\mbox{$q_1, q_2, \ldots, q_N$}; 
they can be written in terms of \mbox{$\{q_{i}\}$}, 
their partial derivatives or 
space/time 
shifts. 
In addition, in the \mbox{$(2+1)$}-dimensional case, 
$f_i$ 
are, in general, nonlocal functions of \mbox{$\{q_{i}\}$} 
involving 
a mixed action of 
\mbox{$\partial_x$}, \mbox{$\partial_x^{-1}$}, and 
\mbox{$\partial_y$,} such as \mbox{$\partial_x^{-1}\partial_y$}. 

Let us start to construct the 
Miura maps from the Lax representation. 
Using the spatial part of the Lax representation (\ref{lin2-1})
that is linear in 
$\Psi_j$, we obtain 
a set of coupled Riccati-type equations
for 
$\Psi_i \Psi_j^{-1}$, i.e., 
\begin{subequations}
\label{Miu1}
\begin{align}
\left( \Psi_i \Psi_j^{-1} \right)_x 
 &= \sum_{k=1}^l U_{ik} \Psi_k \Psi_j^{-1} 
 - \Psi_i \Psi_j^{-1} \sum_{k=1}^l U_{jk} \Psi_k \Psi_j^{-1}
\nonumber \\
 &= U_{ij} + U_{ii} \Psi_i \Psi_j^{-1} 
 - \Psi_i \Psi_j^{-1} U_{jj}
 + \sum_{k (\neq i, j) } U_{ik} \Psi_k \Psi_j^{-1} 
\nonumber \\
& \hphantom{=} \;\,
\mbox{} - \Psi_i \Psi_j^{-1} \sum_{k(\neq j)} U_{jk} \Psi_k \Psi_j^{-1}, 
\hspace{5mm} 1 \le i \neq j \le l
\end{align}
in the continuous-space case (cf.~\cite{Haberman}) and 
%
\begin{align}
& \hphantom{=} \;\, \Psi_{i,n+1} \Psi_{j,n+1}^{\, -1}U_{jj}
 + \Psi_{i,n+1} \Psi_{j,n+1}^{\, -1}\sum_{k(\neq j)}
U_{jk} \Psi_{k,n} \Psi_{j,n}^{\, -1} 
\nonumber \\
 & = U_{ij} + U_{ii} \Psi_{i,n} \Psi_{j,n}^{\, -1}
 + \sum_{k(\neq i,j)} U_{ik} \Psi_{k,n} \Psi_{j,n}^{\, -1}, 
\hspace{5mm} 1 \le i \neq j \le l
\end{align}
\end{subequations}
in the discrete-space case. The latter set of relations 
is 
derived from 
the trivial identity \mbox{$\Psi_{i,n+1}\Psi_{j,n+1}^{-1}
	\Psi_{j,n+1}\Psi_{j,n}^{-1}= \Psi_{i,n+1}\Psi_{j,n}^{-1}$}. 
Here, (\ref{Miu1}) can be viewed as 
a linear algebraic system for the 
matrix elements $U_{ij}$. 
Note that 
not all 
of the 
ratios between different 
components, 
\mbox{$\Psi_i \Psi_j^{-1}$} 
for 
\mbox{$i \neq j$}, 
are 
independent 
quantities. 
We can 
suitably 
choose 
a 
set of \mbox{$l-1$} 
distinct 
\mbox{$\Psi_i \Psi_j^{-1}$} 
\mbox{$(\hspace{1pt}1 \le i \neq j \le l \hspace{1pt})$}
so that all the other \mbox{$\Psi_i \Psi_j^{-1}$} can be expressed 
as a product
of these 
quantities and their inverse. 
In this section, we 
express this 
basis set of \mbox{$\Psi_i \Psi_j^{-1}$} 
as 
\mbox{$\{p_1, p_2, \ldots, p_{l-1} \}$}. 
Typically, 
we set 
\mbox{$ p_j := \Psi_{j+1} \Psi_1^{-1}
$}
or 
\mbox{$ p_j := \Psi_{j+1} \Psi_j^{-1}
$}. 
Thus, (\ref{Miu1}) can be 
reformulated as a system for 
\mbox{$\{p_1, p_2, \ldots, p_{l-1} \}$}, 
their 
$x$-derivatives
or spatial 
shifts, 
and 
\mbox{$\{q_1, q_2, \ldots, q_N\}$}. 
Therefore, it is not ultralocal 
in 
the $p_j$, 
but 
it is ultralocal in the dynamical variables $q_j$ 
because the spatial Lax matrix $U$ is assumed to be 
ultralocal in these variables. 
In this way, 
(\ref{Miu1}) can 
provide 
a (typically) algebraic system 
of \mbox{$l-1$} equations 
for the 
$N$ unknowns $q_j$. 
When 
\mbox{$l-1 >N$}, 
one might consider that 
this system is overdetermined 
and 
has no solution. 
However, this is not the case 
as long as 
the original 
Lax representation 
is consistent and 
provides
a meaningful integrable system 
through the zero-curvature condition. 
In that case, the 
remaining 
\mbox{$l-1-N$} equations 
represent 
relations 
between 
some 
$p_j$, 
for example, 
$p_{j_1}$ is equal to a certain 
differential polynomial of $p_{j_2}$.

Similarly to the spatial part, 
the time part of the Lax representation (\ref{lin2-2}) provides
another set of coupled Riccati-type equations
for $\Psi_i \Psi_j^{-1}$, i.e., 
\begin{subequations}
\label{Miu2}
\begin{align}
\widehat{T}_2 
\left( \Psi_i \Psi_j^{-1} \right)
&= V_{ij} + V_{ii} \Psi_i \Psi_j^{-1} 
 - \Psi_i \Psi_j^{-1} V_{jj}
 + \sum_{k (\neq i, j) } V_{ik} \Psi_k \Psi_j^{-1} 
\nonumber \\
& \hphantom{=} \;\,
\mbox{} - \Psi_i \Psi_j^{-1} \sum_{k(\neq j)} V_{jk} \Psi_k \Psi_j^{-1}, 
\hspace{5mm} 1 \le i \neq j \le l
\end{align}
for the
derivation \mbox{$\widehat{T}_2$}
in the continuous-time case 
and 
\begin{align}
\widehat{T}_2 \left( \Psi_{i} \Psi_{j}^{-1}\right) = 
\left( V_{ij} + \sum_{k(\neq j)} V_{ik} \Psi_{k} \Psi_{j}^{-1} \right)
\left( V_{jj}+ 
\sum_{k(\neq j)} V_{jk} \Psi_{k} \Psi_{j}^{-1} \right)^{-1},  
\nonumber \\
\hspace{5mm} 
1 \le i \neq j \le l
\end{align}
\end{subequations}
for the
shift operator \mbox{$\widehat{T}_2$} 
in the discrete-time case. We can suitably 
take out 
\mbox{$l-1$} 
equations 
from (\ref{Miu2}) and reformulate them as 
the time evolution equations for 
\mbox{$\{p_1, p_2, \ldots, p_{l-1} \}$}. 
Note that 
these equations 
are not ultralocal in 
$q_j$ (cf.~(\ref{NLS-R2}))
and depend on factors such as 
their partial derivatives or
shifts, 
through the 
elements 
of 
the temporal Lax matrix $V$. 

Now, we have all the ingredients needed 
for constructing 
the (inverse) 
Miura maps:
\begin{itemize}
\item[(1)]
$N$ 
evolution equations for the original 
variables 
\mbox{$\{q_1, q_2, \ldots, q_N \}$} 
given by (\ref{evo1}), 
\item[(2)]
(typically algebraic) 
\mbox{$l-1$} 
``ultralocal'' equations
for 
\mbox{$\{q_1, q_2, \ldots, q_N \}$} 
obtained from (\ref{Miu1}), 
wherein the coefficients 
involve 
\mbox{$\{p_1, p_2, \ldots, p_{l-1} \}$},
their
$x$-derivatives
or spatial
shifts,
\item[(3)]
\mbox{$l-1$}
evolution equations for the new 
variables \mbox{$\{p_1, p_2, \ldots, p_{l-1} \}$} 
obtained from (\ref{Miu2}), 
which also 
involve \mbox{$\{q_1, q_2, \ldots, q_N \}$}, 
their partial derivatives, shifts, etc. 
\end{itemize}
%
The key 
observation to be made 
(cf.~(\ref{NLS-R1}))
is that 
we can solve (2) 
to express \mbox{$\min (l-1,N)$} 
of the $N$ 
variables 
\mbox{$\{q_1, q_2, \ldots, q_N \}$} 
in terms of 
the remaining 
\mbox{$\max (N-l+1,0)$} 
$q_j$ 
and 
\mbox{$\min (l-1,N)$} 
of the \mbox{$l-1$} variables 
\mbox{$\{p_1, p_2, \ldots, p_{l-1} \}$}. 
These \mbox{$\min (l-1,N)$} expressions 
define 
a Miura map to the original system of 
$N$ 
evolution equations (1). 
The 
corresponding 
modified system 
is obtained by 
combining the 
\mbox{$\max (N-l+1,0)$} evolution equations 
from (1) 
and the 
\mbox{$\min (l-1,N)$} evolution equations 
from (3), 
wherein the 
unnecessary 
variables 
must 
be eliminated using (2) in order to 
write the modified system in closed form. 
Note that (2) and (3) involve the spectral parameter $\zeta$, 
so that the Miura map and 
the modified system 
also contain
$\zeta$ 
as 
an additional
parameter. 
The (nonstandard) 
Lax representation for 
the modified system 
can be obtained 
directly 
by substituting the defining 
expressions for 
the Miura map into the original Lax representation for (1); 
however, 
the free parameter $\zeta$ 
in the Miura map and the modified system 
must 
be distinguished
from the spectral parameter $\zeta$ in the original Lax representation. 
This is a 
brief sketch of 
our method;
for clarity, we 
discuss the three cases 
\mbox{$l-1 >N$}, \mbox{$l-1 =N$}, and \mbox{$l-1 <N$} 
separately in more detail. 
\begin{itemize}
\item The case of \mbox{$l-1 >N$} 
(Lax representation is ``sparse'' in the dependent variables): 
with a suitable renumbering of the 
\mbox{$p_j$}, 
the Miura map is given by 
\[
(p_1, p_2, \ldots, p_{N}) 
\mapsto (q_1, q_2, \ldots, q_N ).
\]
Thus, 
the 
entire set 
of dependent variables is 
changed. 
%
In solving 
system (2) 
for a general value of $\zeta$, 
we 
need 
to 
identify a basis set 
of 
$N$ new 
variables $p_j$ 
that can be used 
to 
express 
the remaining \mbox{$l-1-N$} 
$p_j$ and the $N$ original variables $q_j$. 
When 
this elimination 
process 
turns out to be too complicated, 
we 
consider 
if 
fixing the 
parameter $\zeta$ at some 
special 
value, say \mbox{$\zeta=0$}, 
can 
simplify 
system (2) 
to yield 
a parameter-less 
Miura map in concise
form. 
%
In that case, 
the Miura map 
and the modified system 
should be written in terms of 
just $N$ new variables $p_j$, 
but 
these $N$ variables 
are not required to 
form a basis 
set 
to 
express the other 
$p_j$ 
explicitly (cf.~\cite{Weiss84,Weiss85}). 
\item The case of \mbox{$l-1 =N$}: 
the Miura map is given by 
\[
(p_{1}, p_{2}, \ldots, p_{l-1}) 
\mapsto (q_1, q_2, \ldots, q_N )
. \]
Thus, 
the 
entire set 
of dependent variables is fully 
replaced
without 
any need to choose 
a subset of the original/new dependent variables; 
conventionally, 
this is the 
case that 
the term ``Miura map'' 
refers to. 
We first 
solve 
system (2) 
to express \mbox{$\{q_1, q_2, \ldots, q_N \}$} 
in terms of 
\mbox{$\{p_1, p_2, \ldots, p_{l-1} \}$}. 
Subsequently,
substituting 
these expressions into (3), 
we obtain 
the modified system 
for the new variables $p_j$.
In the simplest case of \mbox{$l=2$} and \mbox{$N=1$}, 
that is, a \mbox{$2 \times 2$} matrix Lax representation 
containing only one 
dependent variable,
this procedure is 
equivalent to Chen's method~\cite{Chen1,Chen2}. 
Note that the idea of Chen's method 
naturally comes from
the original 
Miura map 
between 
the Gardner equation 
and 
the KdV equation~\cite{Miura68,MGK68} (also see 
Kruskal's article~\cite{Kruskal74}). 
Chen's method 
can be 
suitably modified 
and applied to 
the discrete case~\cite{Chen-JMP1,Yang94}, 
the \mbox{$(2+1)$}-dimensional case~\cite{Chen-JMP2}, 
and the multicomponent case~\cite{Adler08} as well. 
\item The case of \mbox{$l-1 <N$} 
(Lax representation is ``dense'' in the dependent variables):
with a suitable renumbering of the 
\mbox{$q_j$}, 
the Miura map is 
given by 
\[
(q_{1}, q_{2}, \ldots, q_{N-l+1}, p_{1}, p_{2}, \ldots, p_{l-1})
\mapsto (q_1, q_2, \ldots, 
q_N ). 
\]
Thus, only a subset of dependent variables is 
replaced. 
This appears to be 
the most interesting case
and 
we 
concentrate on 
this case in the subsequent sections. 
System (2) of \mbox{$l-1$} equations is 
underdetermined, 
so that we can eliminate only \mbox{$l-1$} of the $N$ original variables 
\mbox{$\{q_1, q_2, \ldots, q_N \}$}, 
which are
relabelled as \mbox{$\{q_{N-l+2}, \ldots, q_{N} \}$} above. 
That is, we 
mix \mbox{$N-l+1$} original variables 
$q_j$ 
with the \mbox{$l-1$} new variables 
$p_j$ to form a closed system; 
actually, 
the way to eliminate \mbox{$l-1$} $q_j$ is, in general, 
nonunique. 
There exist maximally 
$\binom{N}{l-1}$ different ways 
to eliminate \mbox{$l-1$} $q_j$ and mix the original 
and new variables. 
If the original system 
is totally
asymmetric
with respect to 
permutations of the variables \mbox{$\{q_1, q_2, \ldots, q_N \}$}, 
then different eliminations may result in 
entirely different 
modified systems. 
%
Moreover, 
before applying the method, 
we 
can 
consider
any 
invertible 
ultralocal change of the 
dependent variables 
in the original system; 
this change 
further leads to 
distinct
modified systems. 
Therefore, 
we 
need to have an eye for beauty
to sort out 
particularly interesting modified systems. 
\end{itemize}
In the first and second cases, \mbox{$l-1 \ge N$},
our method is 
conceptually similar to the method used in~\cite{Kono90} 
(also see \S3.1 of \cite{AS81}). 
However, 
in contrast to our method, 
the 
method in~\cite{Kono90,AS81} 
uses 
$N$ 
components 
of the linear eigenfunction directly 
as the new dependent variables. 

Before ending 
this section, we remark that the definition 
of 
the 
new 
variables 
\mbox{$\{p_1, p_2, \ldots, p_{l-1} \}$} 
intrinsically allows the following two 
freedoms:
\begin{itemize}
\item
the way 
of choosing
\mbox{$p_k:=\Psi_{i_k}\Psi_{j_k}^{-1}$}, 
which 
are algebraically
independent and 
form 
a basis set 
to express all 
\mbox{$\Psi_i \Psi_j^{-1}$}
\mbox{$(\hspace{1pt}1 \le i \neq j \le l \hspace{1pt})$}, 
is nonunique;
%
\item
for the Lax representation (\ref{lin1}), 
an arbitrary
gauge transformation 
\mbox{$\vt{\psi} \mapsto \vt{\phi}:=g \vt{\psi}$} 
can be considered, 
wherein 
$g$ is a 
constant invertible matrix 
to maintain
the ultralocality of the spatial Lax matrix 
in 
the 
dependent 
variables.
\end{itemize}
%
%
Thus, these two freedoms 
represent a certain group of 
point transformations
that act on 
the set of new dependent variables 
\mbox{$\{p_1, p_2, \ldots, p_{l-1} \}$}. 
In the first case, \mbox{$l-1 >N$}, 
these freedoms may play a critical role in obtaining 
a simple 
modified system in closed form. 
In the second and third cases, \mbox{$l-1 \le N$}, 
they 
are 
not 
as essential as in the first case, 
because 
the entire set \mbox{$\{p_1, p_2, \ldots, p_{l-1} \}$} 
is 
used 
in the modified system.
Nevertheless, 
a good choice 
of the new variables $p_j$ 
is crucial
for obtaining
a highly attractive
modified system. 
%
%
%
%

\section{\mbox{$(1+1)$}-dimensional PDEs}

\label{Exs}
In this section, we apply the 
method 
described in section 2 to 
integrable 
PDEs 
in \mbox{$1+1$} dimensions. 
Four illustrative 
examples, namely,
the Zakharov--Ito system, 
the Jaulent--Miodek system, 
the three-wave interaction system,
and the Yajima--Oikawa system, 
are considered.
%

\subsection{The Zakharov--Ito system}

A two-component generalization of the KdV equation, 
\begin{subequations}
\label{Itoeq}
\begin{align}
& u_t = u_{xxx} + 3uu_x + 3 w w_x,
\\
& w_t = (u w)_x,
\end{align}
\end{subequations}
was proposed 
by 
Ito~\cite{Ito} in 1982.
Actually, 
through the simple 
change of dependent variables 
\mbox{$q:=-u/2$} and \mbox{$r:= -3w^2/16$},
(\ref{Itoeq})
can be rewritten as
\begin{subequations}
\label{ZakIto}
\begin{align}
& q_t = q_{xxx} -6qq_x + 4r_x,
\label{ZakIto-1}
\\
& r_t = -4q_x r - 2qr_x, 
\label{ZakIto-2}
\end{align}
\end{subequations}
which was obtained earlier by Zakharov~\cite{Zakh}.
%
Thus, we may call (\ref{Itoeq}) or (\ref{ZakIto}) 
the Zakharov--Ito system. 
The Lax representation for 
(\ref{ZakIto})
is given by 
the pair of linear equations for a scalar function $\psi$~\cite{Zakh,Boiti83},
\begin{subequations}
\label{ItoLax}
\begin{align}
& \psi_{xx} = \left( \zeta + q + \zeta^{-1} r \right) \psi,
\label{ItoLax-1}
\\
& \psi_t = (4\zeta - 2q)\psi_x +q_x \psi.
\label{ItoLax-2}
\end{align}
\end{subequations}
Indeed, the compatibility condition
$\psi_{xxt} = \psi_{txx}$ for (\ref{ItoLax})
provides (\ref{ZakIto}). 
We introduce a
new variable $v$ as
\mbox{$\psi_{x}/\psi=: v$}.
Then,
the Lax representation (\ref{ItoLax})
implies the pair of relations
\begin{subequations}
\label{ZIpsi}
\begin{align}
& v_{x} +v^2 = \zeta + q + \zeta^{-1} r,
\label{ZIpsi-1}
\\
& v_{t} = \left[ (4\zeta - 2q)v +q_x\right]_x.
\label{ZIpsi-2}
\end{align}
\end{subequations}
Relation (\ref{ZIpsi-1}) can define two different 
Miura maps, 
\mbox{$(q,v) \mapsto (q,r)$} and \mbox{$(v,r) \mapsto (q,r)$}, 
both of which 
retain polynomiality of the 
equations of motion. 

First, we consider the Miura map 
\mbox{$(q,v) \mapsto (q,r)$} with
\mbox{$r = -\zeta^2 + \zeta (-q+v_{x} +v^2)$} 
(cf.~\cite{Liu91} for the \mbox{$\zeta=1$} case). 
Substituting this expression 
for $r$ 
into 
(\ref{ZakIto-1}) 
and combining it with (\ref{ZIpsi-2}), 
we obtain the 
modified system for \mbox{$(q,v)$}, 
\begin{subequations}
\label{mZI1}
\begin{align}
& q_t = q_{xxx} -6qq_x -4 \zeta q_x + 4 \zeta (v_{xx} +2 vv_x),
\label{mZI1-1}
\\
& v_{t} = 4\zeta v_x - 2 (qv)_x +q_{xx}.
\label{mZI1-2}
\end{align}
\end{subequations}
Note that 
the 
linear terms 
with the 
first-order 
$x$-differentiation
can be eliminated using
a Galilean transformation. 
Indeed, by setting 
\mbox{$q=: q'-2\zeta$}, 
\mbox{$\partial_t - 8 \zeta\partial_x =: \partial_{t'}$}, 
and \mbox{$\partial_x =: \partial_{x'}$}, 
(\ref{mZI1})
is simplified 
to 
\begin{subequations}
\label{}
\begin{align}
& q_t = q_{xxx} -6qq_x + 4 \zeta (v_{xx} +2 vv_x),
\label{}
\\
& v_{t} = q_{xx}- 2 (qv)_x,
\label{}
\end{align}
\end{subequations}
where the prime 
is omitted for brevity. 

Second, we consider the Miura map 
\mbox{$(v,r) \mapsto (q,r)$} 
with \mbox{$q=-\zeta- \zeta^{-1} r + v_{x} +v^2$}.
Substituting this expression for $q$ 
into (\ref{ZIpsi-2}) and (\ref{ZakIto-2}), 
we obtain the modified system for \mbox{$(v,r)$}, 
\begin{align}
& v_{t} = 6 \zeta v_x +v_{xxx} -6v^2 v_x 
	+ 2 \zeta^{-1}(vr)_x -\zeta^{-1} r_{xx},
\nonumber \\
& r_t = 2\zeta r_x +6\zeta^{-1} rr_x -4 (v_x+ v^2)_x r - 2(v_x+ v^2) r_x.
\nonumber 
\end{align}
%
By setting \mbox{$r=: \zeta p^2$}, we can rewrite 
this system 
in the form
\begin{subequations}
\label{4.57}
\begin{align}
& v_{t} = 6 \zeta v_x +v_{xxx} -6v^2 v_x 
	+ 2 (vp^2)_x - (p^2)_{xx},
\\
& p_t = 2\zeta p_x +6 p^2 p_x -2 (v_x p + v^2 p)_x.
\end{align}
\end{subequations}
This 
two-component mKdV system, 
as well as 
the Miura map 
to the Zakharov--Ito system, 
is 
presented in section~4.2.17 of \cite{TW05}, 
wherein one can also find information on the \mbox{$\zeta \to 0$} limit.
A rather similar system is 
studied in~\cite{Liu91}.

\subsection{The Jaulent--Miodek system}

Another two-component generalization of the KdV hierarchy 
was proposed by Jaulent and Miodek~\cite{JM}. 
Its first nontrivial flow 
is
\begin{subequations}
\label{JM1st}
\begin{align}
& q_t= r_{xxx} + 2 q_x r +4qr_x,
\label{JM-1}
\\
& r_t= 4q_x +6rr_x, 
\label{JM-2}
\end{align}
\end{subequations}
while the 
next 
flow allows
the reduction \mbox{$r=0$} to the KdV equation~\cite{JM}. 
The Lax representation for
(\ref{JM1st})
is given by
the pair of linear equations,
%
\begin{subequations}
\label{JMLax}
\begin{align}
& \psi_{xx} + \left( q + \zeta r \right) \psi = \zeta^{2}\psi,
\label{JMLax-1}
\\
& \psi_t = (4\zeta + 2r)\psi_x -r_x \psi.
\label{JMLax-2}
\end{align}
\end{subequations}
We introduce a 
new variable $u$ as
\mbox{$\psi_{x}/\psi=: u$}.
Thus,
the Lax representation (\ref{JMLax})
implies the pair of relations
\begin{subequations}
\label{JMpsi}
\begin{align}
& u_{x} +u^2 +q + \zeta r= \zeta^2,
\label{JMpsi-1}
\\
& u_{t} = \left[ (4\zeta + 2r)u -r_x\right]_x.
\label{JMpsi-2}
\end{align}
\end{subequations}
Relation (\ref{JMpsi-1}) can define two different
Miura maps to the Jaulent--Miodek system (\ref{JM1st}), i.e.,
\mbox{$(q,u) \mapsto (q,r)$} and \mbox{$(u,r) \mapsto (q,r)$}; 
both of them 
retain polynomiality of the
equations of motion.

First, we consider the Miura map
\mbox{$(q,u) \mapsto (q,r)$} with
\mbox{$r = \zeta -
\zeta^{-1}(q+u_{x} +u^2)$}. 
Substituting this expression
for $r$
into
(\ref{JM-1}) and (\ref{JMpsi-2}) 
and rescaling the time variable 
as 
\mbox{$\partial_t =: -\zeta^{-1} \partial_{\hat{t}}$},
we obtain the
modified system for \mbox{$(q,u)$},
\begin{subequations}
\label{mJM2}
\begin{align}
& q_{\hat{t}} = -2\zeta^2 q_x + q_{xxx} +6qq_x 
+u_{xxxx} + (u^2)_{xxx} 
\nonumber \\
& \hspace{7.5mm} \mbox{}
+2 q_x u_x +4 q u_{xx} +2q_x u^2 + 8 q uu_x,
\label{mJM2-1}
\\[1.5mm]
& u_{\hat{t}} = -6\zeta^2 u_x -u_{xxx} +6 u^2 u_x -q_{xx} +2(qu)_x.
\label{mJM2-2}
\end{align}
\end{subequations}
Interestingly, 
(\ref{mJM2})
resembles
a system of the coupled KdV--mKdV type. 

Second, we consider the Miura map
\mbox{$(u,r) \mapsto (q,r)$} with
\mbox{$q = \zeta^2 -\zeta r -u_{x} -u^2$}
(cf.~\cite{Alonso} for 
\mbox{$\zeta=0$}
and~\cite{Zeng98} for general $\zeta$).
Substituting this expression for $q$
into (\ref{JM-2}) 
and combining it with (\ref{JMpsi-2}),
we obtain the
modified system for \mbox{$(u,r)$},
\begin{subequations}
\label{mJM1}
\begin{align}
& u_t= 4\zeta u_x -r_{xx} + 2(ur)_x, 
\\
& r_t= -4\zeta r_x - 4u_{xx} - 8u u_x + 6rr_x.
\end{align}
\end{subequations}
Note that 
the parameter $\zeta$ in (\ref{mJM1}) 
is nonessential
and can be set equal to zero 
by the Galilean transformation 
\mbox{$r=: r'+2\zeta$},
\mbox{$\partial_t - 8 \zeta\partial_x =: \partial_{t'}$},
\mbox{$\partial_x =: \partial_{x'}$}. 
Thus, (\ref{mJM1}) is 
equivalent to 
the first nontrivial flow of the modified Jaulent--Miodek hierarchy 
studied in~\cite{Alonso}.

\subsection{The three-wave interaction system}


We consider 
the 
three-wave interaction system~\cite{ZM73}
in the 
nonreduced form: 
\begin{subequations}
\label{3W}
\begin{align}
& u_{1,t} -c_{1} u_{1,x} + (c_{2}-c_{3})u_3 v_2 =0, 
\label{3W-1} \\
& u_{2,t} -c_{2} u_{2,x} + (c_{3}-c_{1})u_3 v_1 =0, 
\label{3W-2} \\
& u_{3,t} -c_{3} u_{3,x} + (c_{2}-c_{1})u_1 u_2 =0, 
\label{3W-3} \\
& v_{1,t} -c_{1} v_{1,x} + (c_{3}-c_{2})v_3 u_2 =0, 
\label{3W-4} \\
& v_{2,t} -c_{2} v_{2,x} + (c_{1}-c_{3})v_3 u_1 =0, 
\label{3W-5} \\
& v_{3,t} -c_{3} v_{3,x} + (c_{1}-c_{2})v_1 v_2 =0.
\label{3W-6} 
\end{align}
\end{subequations}
Here, the 
constants $c_j$ 
\mbox{$(j=1,2,3)$}
are 
parametrized 
in terms of the 
constants $\alpha_j$ and $\beta_j$ as 
\begin{equation}
c_1= \frac{\beta_1-\beta_2}{\alpha_1-\alpha_2}, \hspace{5mm}
c_2= \frac{\beta_2-\beta_3}{\alpha_2-\alpha_3}, \hspace{5mm}
c_3= \frac{\beta_3-\beta_1}{\alpha_3-\alpha_1}.
\label{c-a-b}
\end{equation}
Thus, they can be considered the ``slopes'' 
of the three sides of 
a triangle. 
Specific 
integrable 
systems describing 
triad wave interactions 
are derived 
by imposing a 
complex conjugacy 
reduction on
(\ref{3W}) 
to halve the number of 
dependent variables~\cite{ZS3,AH75}. 
%
%
%
The Lax representation for the three-wave interaction system 
(\ref{3W}) with (\ref{c-a-b}) is given by 
the pair of linear PDEs~\cite{AH75},
\begin{subequations}
\label{3W-UV}
\begin{align}
& \left[
\begin{array}{c}
 \Psi_1 \\
 \Psi_2 \\
 \Psi_3 \\
\end{array}
\right]_x
= 
\left[
\begin{array}{ccc}
\mathrm{i}\alpha_1\zeta & u_1 & u_3 \\
 v_1 & \mathrm{i}\alpha_2\zeta & u_2 \\
 v_3 & v_2 & \mathrm{i}\alpha_3\zeta \\
\end{array}
\right]
\left[
\begin{array}{c}
 \Psi_1 \\
 \Psi_2 \\
 \Psi_3 \\
\end{array}
\right],
\label{3W-U}
\\[2mm]
& \left[
\begin{array}{c}
 \Psi_1 \\
 \Psi_2 \\
 \Psi_3 \\
\end{array}
\right]_t
= 
\left[
\begin{array}{ccc}
\mathrm{i}\beta_1\zeta & c_1 u_1 & c_3 u_3 \\
 c_1 v_1 & \mathrm{i}\beta_2\zeta & c_2 u_2 \\
 c_3 v_3 & c_2 v_2 & \mathrm{i}\beta_3\zeta \\
\end{array}
\right]
\left[
\begin{array}{c}
 \Psi_1 \\
 \Psi_2 \\
 \Psi_3 \\
\end{array}
\right].
\label{3W-V}
\end{align}
\end{subequations}
%
%
%
With
\mbox{$\Psi_2/\Psi_1 =: w_1$} and \mbox{$\Psi_3/\Psi_1 =: w_3$},
(\ref{3W-U}) and (\ref{3W-V}) imply 
\begin{subequations}
\label{3W-x}
\begin{align}
& w_{1,x} = v_1 + \mathrm{i}(\alpha_2-\alpha_1) \zeta w_1 + u_2 w_3 
	-(u_1 w_1 + u_3 w_3)w_1,
\label{}
\\
& w_{3,x} = v_3 + \mathrm{i}(\alpha_3-\alpha_1) \zeta w_3 + v_2 w_1 
	-(u_1 w_1 + u_3 w_3)w_3,
\label{}
\end{align}
\end{subequations}
and 
\begin{subequations}
\label{3W-t}
\begin{align}
& w_{1,t} = c_1 v_1 + \mathrm{i}(\beta_2-\beta_1) \zeta w_1 + c_2 u_2 w_3 
	-(c_1 u_1 w_1 + c_3 u_3 w_3)w_1,
\label{}
\\
& w_{3,t} = c_3 v_3 + \mathrm{i}(\beta_3-\beta_1) \zeta w_3 + c_2 v_2 w_1 
	-(c_1 u_1 w_1 + c_3 u_3 w_3)w_3,
\label{}
\end{align}
\end{subequations}
respectively. 
Thus, 
relations (\ref{3W-x}) can 
define the Miura map 
\mbox{$(u_1,u_2,u_3,w_1,v_2,w_3)$}
\\
\mbox{$\mapsto (u_1,u_2,u_3,v_1,v_2,v_3)$}, 
which replaces two of the six dependent variables. 
Using (\ref{3W-x}), we can eliminate $v_1$ and $v_3$ in 
(\ref{3W-t})
to obtain 
\begin{subequations}
\label{m3W1-1}
\begin{align}
& w_{1,t} - c_1 w_{1,x} + (c_1 -c_2) u_2 w_3 -(c_1 - c_3 )u_3 w_3 w_1=0,
\label{m3W1-11}
\\
& w_{3,t} - c_3 w_{3,x} + (c_3 -c_2) v_2 w_1 -(c_3-c_1) u_1 w_1 w_3=0.
\label{m3W1-12}
\end{align}
\end{subequations}
In the same manner, 
(\ref{3W-2}) and (\ref{3W-5}) can be rewritten as 
\begin{subequations}
\label{m3W1-2}
\begin{align}
& u_{2,t} -c_{2} u_{2,x} + (c_{3}-c_{1})u_3 
 \left[ w_{1,x} - \mathrm{i}(\alpha_2-\alpha_1) \zeta w_1 - u_2 w_3 
	+(u_1 w_1 + u_3 w_3)w_1 \right]=0,
\label{m3W1-21} \\
& v_{2,t} -c_{2} v_{2,x} + (c_{1}-c_{3})
 \left[ w_{3,x} - \mathrm{i}(\alpha_3-\alpha_1) \zeta w_3 - v_2 w_1 
	+(u_1 w_1 + u_3 w_3)w_3\right] u_1 =0.
\label{m3W1-22} 
\end{align}
\end{subequations}
The six equations, 
(\ref{3W-1}), (\ref{3W-3}), (\ref{m3W1-1}), and (\ref{m3W1-2}), 
comprise the 
modified system for 
\mbox{$(u_1,u_2,u_3,w_1,v_2,w_3)$} 
in closed form. 
However, the system 
appears
asymmetric 
and 
does not allow 
a 
complex conjugacy 
reduction directly. 
To 
rewrite the modified system in a more symmetric 
form, 
we 
apply a 
point transformation 
to 
some of the variables that 
were unchanged 
in the 
Miura map. 

Comparing 
(\ref{3W-1}) and (\ref{3W-3}) with (\ref{m3W1-11}) and (\ref{m3W1-12}), 
respectively, 
we 
can naturally move from 
the old variable $v_2$ to 
a new variable 
$w_2$ as
\begin{equation}
(c_2-c_3) v_2 =: (c_2-c_1)w_2 + (c_1-c_3)u_1 w_3. 
\label{v_2-w_2}
\end{equation}
Here, we assume that $c_1$, $c_2$, and $c_3$ are 
pairwise distinct. 
Noting the identity \mbox{$(\alpha_1-\alpha_2)(c_2-c_1)
	=(\alpha_1-\alpha_3)(c_2-c_3)$}, 
we can 
express the modified system for \mbox{$(u_1,u_2,u_3,w_1,w_2,w_3)$} 
in the desired form, 
\begin{subequations}
\label{m3W2}
\begin{align}
& u_{1,t} -c_{1} u_{1,x} + (c_{2}-c_{1})u_3 w_2 
	+ (c_{1}-c_{3})u_1 u_3 w_3 =0,
\label{m3W2-1} \\
& u_{2,t} -c_{2} u_{2,x} + (c_{3}-c_{1})u_3 w_{1,x} 
  +\mathrm{i}\zeta (\alpha_1 -\alpha_2 ) (c_3-c_1) u_3 w_1 
\nonumber \\[-1mm]
&\hspace{6mm}-(c_3-c_1) u_2 u_3 w_3 +(c_3-c_1)(u_1 w_1 +u_3 w_3 ) u_3 w_1 =0,
\label{m3W2-2} \\
& u_{3,t} -c_{3} u_{3,x} + (c_{2}-c_{1})u_1 u_2 =0,
\label{m3W2-3} \\
& w_{1,t} -c_{1} w_{1,x} + (c_{1}-c_{2})w_3 u_2 
+(c_{3}-c_{1})w_1 w_3 u_3 
=0,
\label{m3W2-4} \\
& w_{2,t} -c_{2} w_{2,x} - (c_{1}-c_{3})w_3 u_{1,x} 
  +\mathrm{i}\zeta (\alpha_1 -\alpha_2 ) (c_1-c_3) w_3 u_1 
\nonumber \\[-1mm]
&\hspace{6mm}-(c_1-c_3) w_2 w_3 u_3 +(c_1-c_3)(u_1 w_1 +u_3 w_3 ) w_3 u_1 =0,
\label{m3W2-5} \\
& w_{3,t} -c_{3} w_{3,x} + (c_{1}-c_{2})w_1 w_2 =0.
\label{m3W2-6}
\end{align}
\end{subequations}
The 
``refined'' Miura map 
\mbox{$(u_1,u_2,u_3,w_1,w_2,w_3) \mapsto (u_1,u_2,u_3,v_1,v_2,v_3)$} 
from (\ref{m3W2}) to (\ref{3W}) is 
given by combining 
(\ref{3W-x}) and (\ref{v_2-w_2}). 
Actually, 
it is possible to eliminate the quartic terms in 
(\ref{m3W2-2}) and (\ref{m3W2-5}) 
by further applying a nonlocal transformation. 
Indeed, the simplest conservation law 
for (\ref{m3W2}), 
\[
( u_1 w_1 + u_3 w_3 )_t = ( c_1 u_1 w_1 + c_3 u_3 w_3 )_x, 
\]
motivates us to introduce 
the new set of variables 
\mbox{$(q_1,q_2,q_3,r_1,r_2,r_3)$} by
\begin{align}
& u_1=: q_1 \mathrm{e}^{\int^{x} ( u_1 w_1 + u_3 w_3 )\mathrm{d}x'}, 
\hspace{5mm}
u_2=: q_2 \mathrm{e}^{\delta \int^{x} ( u_1 w_1 + u_3 w_3 )\mathrm{d}x'}, 
\hspace{5mm}
u_3=: q_3 \mathrm{e}^{(1+ \delta) \int^{x} ( u_1 w_1 + u_3 w_3 )\mathrm{d}x'}, 
\nonumber \\
& w_1=: r_1 \mathrm{e}^{-\int^{x} ( u_1 w_1 + u_3 w_3 )\mathrm{d}x'}, 
\hspace{5mm}
w_2=: r_2 \mathrm{e}^{-\delta \int^{x} ( u_1 w_1 + u_3 w_3 )\mathrm{d}x'}, 
\hspace{5mm}
w_3=: r_3 \mathrm{e}^{-(1+ \delta) \int^{x} ( u_1 w_1 + u_3 w_3 )
	\mathrm{d}x'}.
\label{nonlo}
\end{align}
Here, $\delta$ is an arbitrary constant. 
Note that 
\mbox{$\exp \left[ \int^{x} ( u_1 w_1 + u_3 w_3 )\mathrm{d}x'
\right]$} can be identified with 
\mbox{$\Psi_1 \exp (-\mathrm{i}\alpha_1 \zeta x - \mathrm{i}\beta_1 \zeta t)$} 
in the Lax representation (\ref{3W-UV}), up to a 
multiplicative constant. 
Under suitable boundary conditions, the 
modified system (\ref{m3W2}) is transformed to the 
form
\begin{subequations}
\label{m3W3}
\begin{align}
& q_{1,t} -c_{1} q_{1,x} + (c_{2}-c_{1})q_3 r_2 =0,
\label{m3W3-1} \\
& q_{2,t} -c_{2} q_{2,x} + (c_{3}-c_{1})q_3 r_{1,x} 
  +\mathrm{i}\zeta (\alpha_1 -\alpha_2 ) (c_3-c_1) q_3 r_1 
\nonumber \\[-1mm]
&\hspace{6mm}+\delta(c_1-c_2) q_1 q_2 r_1 
+\left[ (c_1-c_3) + \delta(c_3-c_2) \right] q_2 q_3 r_3 
 =0,
\label{m3W3-2} \\
& q_{3,t} -c_{3} q_{3,x} + (c_{2}-c_{1})q_1 q_2 
+ (1+\delta) (c_{1}-c_{3})q_1 q_3 r_1 =0,
\label{m3W3-3} \\
& r_{1,t} -c_{1} r_{1,x} + (c_{1}-c_{2})r_3 q_2 =0,
\label{m3W3-4} \\
& r_{2,t} -c_{2} r_{2,x} - (c_{1}-c_{3})r_3 q_{1,x} 
  +\mathrm{i}\zeta (\alpha_1 -\alpha_2 ) (c_1-c_3) r_3 q_1 
\nonumber \\[-1mm]
&\hspace{6mm} 
+\delta (c_2-c_1) r_1 r_2 q_1 
+\left[ (c_3-c_1) + \delta (c_2-c_3) \right] r_2 r_3 q_3 
=0,
\label{m3W3-5} \\
& r_{3,t} -c_{3} r_{3,x} + (c_{1}-c_{2})r_1 r_2 
+ (1+\delta) (c_{3}-c_{1})r_1 r_3 q_1=0.
\label{m3W3-6}
\end{align}
\end{subequations}
Some special cases, such as \mbox{$\delta=-1 \; \mathrm{or} \; 0$} 
and \mbox{$\zeta=0$}, 
appear to be particularly interesting. 

Note that 
(\ref{v_2-w_2}) is not the only point transformation 
that can convert 
the modified system for 
\mbox{$(u_1,u_2,u_3,w_1,v_2,w_3)$} to a symmetric form 
that allows
a complex conjugacy reduction. 
Let us consider the simplest case of \mbox{$\zeta=0$} 
and change the variables $u_2$ and $v_2$ to 
$\widehat{u}_2$ and $\widehat{w}_2$ as 
\begin{equation}
u_2 -u_3 w_1 =:  \widehat{u}_2, 
%
\hspace{5mm} v_2 -u_1 w_3 =: 
	\frac{\alpha_1-\alpha_3}{\alpha_1-\alpha_2}\widehat{w}_2.
\label{u_2-v_2}
\end{equation}
Thus, the modified system for 
\mbox{$(u_1,\widehat{u}_2,u_3,w_1,\widehat{w}_2,w_3)$} 
can be 
written 
as 
\begin{subequations}
\label{m3W4}
\begin{align}
& u_{1,t} -c_{1} u_{1,x} + (c_{2}-c_{1})u_3 \widehat{w}_2 
	+ (c_{2}-c_{3})u_1 u_3 w_3 =0,
\label{m3W4-1} \\ 
& \widehat{u}_{2,t} -c_{2} \widehat{u}_{2,x} + (c_{1}-c_{2})
	u_1 \widehat{u}_2 w_1 +(c_2-c_3) \widehat{u}_2 u_3 w_3 
\nonumber \\[-1mm]
&\hspace{6mm}
+(c_3-c_2) \left[ (u_3 w_1)_x + (u_1 w_1 - u_3 w_3 )u_3 w_1 \right]
=0,
\label{m3W4-2} \\
& u_{3,t} -c_{3} u_{3,x} + (c_{2}-c_{1})u_1 \widehat{u}_2 
	+ (c_{2}-c_{1})u_1 u_3 w_1 =0,
\label{m3W4-3} \\
& w_{1,t} -c_{1} w_{1,x} + (c_{1}-c_{2})w_3 \widehat{u}_2 
+(c_{3}-c_{2})w_1 w_3 u_3 =0,
\label{m3W4-4} \\
& \widehat{w}_{2,t} -c_{2} \widehat{w}_{2,x} + (c_{2}-c_{1})
	w_1 \widehat{w}_2 u_1 + (c_3-c_2) \widehat{w}_2 w_3 u_3 
\nonumber \\[-1mm]
&\hspace{6mm}
+(c_3-c_2) \left[ (w_3 u_1)_x - (u_1 w_1 - u_3 w_3 )w_3 u_1 \right]
=0,
\label{m3W4-5} \\
& w_{3,t} -c_{3} w_{3,x} + (c_{1}-c_{2})w_1 \widehat{w}_2 
	+(c_{1}-c_{2})w_1 w_3 u_1 =0.
\label{m3W4-6}
\end{align}
\end{subequations}
%
The 
associated spectral problem 
in the 
canonical form 
can be obtained 
by applying a 
gauge transformation to 
(\ref{3W-U}) with (\ref{3W-x}) and (\ref{u_2-v_2}). 
It is given by 
\begin{align}
& \left[
\begin{array}{c}
 \Phi_1 \\
 \Phi_2 \\
 \Phi_3 \\
\end{array}
\right]_x
=
\left[
\begin{array}{ccc}
\mathrm{i}\alpha_1\zeta + u_1 w_1 +u_3 w_3
	& (\alpha_2-\alpha_1)u_1
	& (\alpha_3-\alpha_1)u_3 \\
\mathrm{i}\zeta w_1 & \mathrm{i}\alpha_2\zeta -u_1 w_1& 
	\frac{\alpha_3-\alpha_1}{\alpha_2-\alpha_1}\widehat{u}_2 \\
\mathrm{i}\zeta w_3 & \widehat{w}_2 & \mathrm{i}\alpha_3\zeta -u_3 w_3 \\
\end{array}
\right]
\left[
\begin{array}{c}
 \Phi_1 \\
 \Phi_2 \\
 \Phi_3 \\
\end{array}
\right].
\label{3W-U'}
\end{align}
Thus, 
(\ref{u_2-v_2}) is 
a 
natural point transformation 
from the point of view of a Lax representation. 
In the same way as for (\ref{m3W2}), 
we can also consider a nonlocal transformation like (\ref{nonlo}). 
Indeed, 
using 
\begin{align}
& u_1\mathrm{e}^{-\gamma\int^{x} ( u_1 w_1 + u_3 w_3 )\mathrm{d}x'},
\hspace{5mm}
 \widehat{u}_2 \mathrm{e}^{-\delta \int^{x} ( u_1 w_1 + u_3 w_3 )\mathrm{d}x'},
\hspace{5mm}
 u_3 \mathrm{e}^{-(\gamma+ \delta) \int^{x} ( u_1 w_1 + u_3 w_3 )\mathrm{d}x'},
\nonumber \\
& w_1\mathrm{e}^{\gamma\int^{x} ( u_1 w_1 + u_3 w_3 )\mathrm{d}x'},
\hspace{5mm}
 \widehat{w}_2\mathrm{e}^{\delta \int^{x} ( u_1 w_1 + u_3 w_3 )\mathrm{d}x'},
\hspace{5mm}
 w_3 \mathrm{e}^{(\gamma+ \delta) \int^{x} ( u_1 w_1 + u_3 w_3 )
        \mathrm{d}x'}
\nonumber
\end{align}
as the new set of variables 
and 
choosing the parameters $\gamma$ and $\delta$ 
appropriately, 
we 
obtain 
a 
simplified 
version of the modified system (\ref{m3W4}). 
In particular, it is possible to eliminate
the cubic terms 
in (\ref{m3W4-1}), (\ref{m3W4-3}), (\ref{m3W4-4}), and 
(\ref{m3W4-6}). 

\subsection{The Yajima--Oikawa system}

We consider the Yajima--Oikawa system~\cite{YO} 
written 
in the general 
nonreduced form, 
\begin{subequations}
\label{YOsys}
\begin{align}
& \mathrm{i} Q_{t} + Q_{xx} - PQ =O, 
\label{YO-1} \\
& \mathrm{i} R_{t} - R_{xx} + RP =O, 
\label{YO-2} \\
& \mathrm{i} P_{t} + 2 (QR)_x = O.
\label{YO-3} 
\end{align}
\end{subequations}
%
The Lax representation for 
(\ref{YOsys}) is given by 
the set of linear equations (cf.~\cite{Ab78}), 
\begin{subequations}
\label{YO-Lax}
\begin{align}
& \Psi_{xx} =\zeta \Psi + P\Psi + Q\Phi, 
\label{YO-L1} \\
& \Phi_{x} =R\Psi, 
\label{YO-L2} \\
& \mathrm{i} \Psi_{t} + Q\Phi = O,
\label{YO-L3} \\
& \mathrm{i} \Phi_{t} + R\Psi_x -R_x \Psi -\zeta \Phi = O.
\label{YO-L4} 
\end{align}
\end{subequations}
Because 
the linear eigenfunction comprises 
two components $\Psi$ and $\Phi$, 
there exist two 
distinct
ways to define a 
Miura map to (\ref{YOsys}). 

First, we set 
\mbox{$\Phi \Psi^{-1} =: \widehat{R}$}
and \mbox{$\Psi_x \Psi^{-1} =: \widehat{P}$}.
Thus, 
the spatial part of the Lax representation, 
(\ref{YO-L2}) and (\ref{YO-L1}), 
implies that 
\[
R= \widehat{R}_x + \widehat{R}\widehat{P}, 
\hspace{5mm}
P= -\zeta I + \widehat{P}_x +\widehat{P}^2 -Q\widehat{R}. 
\]
%
These relations define 
the Miura map \mbox{$(Q,\widehat{R},\widehat{P}) \mapsto (Q,R,P)$}.
Using (\ref{YO-L1}) and (\ref{YO-L2}), 
the time part of the Lax representation, 
(\ref{YO-L3}) and (\ref{YO-L4}), 
provides the 
evolution equations for 
$\widehat{R}$ and $\widehat{P}$.
Combining 
them with (\ref{YO-1}), we arrive at 
the 
modified 
system for \mbox{$(Q,\widehat{R},\widehat{P})$}, 
%
\begin{subequations}
\label{mYO1sys}
\begin{align}
& \mathrm{i} Q_{t}+\zeta Q + Q_{xx} -\widehat{P}_x Q -\widehat{P}^2 Q 
	+Q\widehat{R}Q=O, 
\label{mYO1-1} \\
& \mathrm{i} \widehat{R}_{t}-\zeta \widehat{R} - \widehat{R}_{xx} 
	-\widehat{R}\widehat{P}_x 
	+\widehat{R}\widehat{P}^2- \widehat{R}Q\widehat{R} =O, 
\label{mYO1-2} \\
& \mathrm{i} \widehat{P}_{t} + (Q\widehat{R})_x +Q\widehat{R}\widehat{P} 
	-\widehat{P}Q\widehat{R}= O.
\label{mYO1-3} 
\end{align}
\end{subequations}
%
In the case of scalar variables, 
this system with \mbox{$\zeta=0$}
was studied in~\cite{New78,New79}, 
and 
the Miura map 
to the Yajima--Oikawa system
(\ref{YOsys}) 
is 
also known~\cite{Liu94}. 
Note that the parameter $\zeta$ is 
nonessential 
if we consider the modified system (\ref{mYO1sys}) separately. 

Moreover, 
we can 
transform (\ref{mYO1sys}) to a simpler form. 
Indeed, the pair of relations \mbox{$\Psi_x = \widehat{P} \Psi$} and 
\mbox{$\mathrm{i}\Psi_t = -Q \widehat{R} \Psi$} motivates us to 
introduce the new 
set of variables as
\[
q:= \Psi^{-1} Q, \hspace{5mm} r:= \widehat{R} \Psi = \Phi, 
\hspace{5mm} p:=\Psi^{-1} \widehat{P} \Psi = \Psi^{-1} \Psi_x .
\]
Then, 
it is easy to show that 
they satisfy the closed 
system, 
\begin{align}
& \mathrm{i} q_{t}+\zeta q + q_{xx} +2 pq_x =O,  
\nonumber \\
& \mathrm{i} r_{t}-\zeta r - r_{xx} +2 r_x p=O,  
\nonumber \\
& \mathrm{i} p_{t} +(qr)_x +pqr -qrp =O, 
\nonumber 
\end{align}
which could be called the derivative Yajima--Oikawa system. 
Again, 
the scalar case with \mbox{$\zeta=0$} 
was 
studied in~\cite{New78,New79}. 

Second, we set 
\mbox{$\Psi \Phi^{-1} (=\widehat{R}^{\hspace{1pt}-1})=: \widehat{Q}$}. 
Because (\ref{YO-L2}) implies 
the relation 
\mbox{$\Phi_x \Phi^{-1} =R \widehat{Q}$}, 
(\ref{YO-L1}) 
gives 
\begin{equation}
Q= \widehat{Q}_{xx} + 2 \widehat{Q}_x R \widehat{Q}
	+ \widehat{Q} \left( R \widehat{Q} \right)_x
	+ \widehat{Q} R \widehat{Q} R \widehat{Q} 
	-\zeta \widehat{Q} -P\widehat{Q}. 
\label{YO-Miura2}
\end{equation}
This relation defines the Miura map 
\mbox{$(\widehat{Q},R,P) \mapsto (Q,R,P)$}, 
which changes only one of the three variables. 
Using also the time part of the Lax representation, 
(\ref{YO-L3}) and (\ref{YO-L4}), we 
obtain 
the evolution equation 
for $\widehat{Q}$, 
\begin{equation}
\mathrm{i}\widehat{Q}_t +\widehat{Q}_{xx} 
+ 2 \left( \widehat{Q} R \right)_x \widehat{Q} -P\widehat{Q} =O.
\label{mYOpre1}
\end{equation}
Substituting (\ref{YO-Miura2}) into 
(\ref{YO-3}), 
we obtain the evolution equation for $P$, 
\begin{align}
& \mathrm{i} P_t -2\zeta \left( \widehat{Q} R\right)_{x} 
 + 2\left( \widehat{Q}_{xx} R \right)_x 
\nonumber \\
& 
 + \left[ 4\widehat{Q}_x R\widehat{Q}R 
+2\widehat{Q} \left( R \widehat{Q} \right)_x R 
-2P \widehat{Q} R +2\left( \widehat{Q}R \right)^3 \right]_x=O.
\label{mYOpre2}
\end{align}
%
The three equations 
(\ref{mYOpre1}), 
(\ref{YO-2}), and 
(\ref{mYOpre2}) comprise the 
modified 
system 
for 
\mbox{$(\widehat{Q},R,P)$}. 
However, 
this 
modified system 
has no 
physical significance in its present form, because 
it looks 
asymmetric 
with respect to $\widehat{Q}$ and $R$. 
That is, 
we 
cannot directly 
impose 
a conjugate relation between 
the two 
variables. 
To restore 
a desired symmetry, 
we only have to change the 
variable $P$ as 
\mbox{$P-2\widehat{Q}_x R +\zeta I =:\widetilde{P}$}, where 
the tilde does not 
denote
the forward shift in the discrete-time case. 
Thus, the modified system 
takes the form
\begin{subequations}
\label{mYO2sys}
\begin{align}
& \mathrm{i}\widehat{Q}_t +\zeta \widehat{Q} +\widehat{Q}_{xx} 
	+ 2 \widehat{Q} R_x \widehat{Q} -\widetilde{P} \widehat{Q} =O,
\label{mYO2-1}
\\[1mm]
& \mathrm{i} R_{t} -\zeta R - R_{xx} 
	+2R\widehat{Q}_xR  + R\widetilde{P} =O,
\label{mYO2-2}
\\[1mm]
& \mathrm{i} \widetilde{P}_t +2\left( \widehat{Q}_x R_x \right)_{x} 
 - 2\left( \widehat{Q} R_x \right)_x \widehat{Q} R 
 + 2\widehat{Q} R \left( \widehat{Q}_x R \right)_x 
 - 2 \widehat{Q}_x R \widehat{Q}_x R  + 2 \widehat{Q} R_x \widehat{Q} R_x
\nonumber \\[-0.5mm]
& -2 \widetilde{P} \widehat{Q} R_x
 -2\widehat{Q}_x R \widetilde{P}
 +2\left[ \left( \widehat{Q}R \right)^3 \right]_x =O.
\label{mYO2-3}
\end{align}
\end{subequations}
The Miura map to the original 
Yajima--Oikawa 
system (\ref{YOsys}) 
is given
by 
the relations 
\[
Q= \widehat{Q}_{xx} 
        + \widehat{Q} \left( R \widehat{Q} \right)_x
        + \widehat{Q} R \widehat{Q} R \widehat{Q}
        -\widetilde{P}\widehat{Q}, 
\hspace{5mm}
P= -\zeta I + 2\widehat{Q}_x R + \widetilde{P}.
\]
Note that 
by setting 
\mbox{$\widetilde{P}= P' + \alpha 
(\widehat{Q}_x R -\widehat{Q} R_x ) + \beta (\widehat{Q}R)^2$}, we 
obtain a two-parameter deformation of 
the modified system (\ref{mYO2sys}).

\section{Differential-difference equations}

In this section, we apply 
our 
method to 
differential-difference equations 
in \mbox{$1+1$} dimensions, 
such as the 
Toda lattice, 
the Belov--Chaltikian lattice, 
the relativistic Toda lattice, 
and the Ablowitz--Ladik lattice, 
wherein 
the spatial variable $n$ takes discrete values. 
We can also 
apply our method in the fully 
discrete case, wherein 
both the space and 
time variables 
are discretized. 
%
However, the 
results 
appear to be 
too 
complicated compared with the continuous-time case, 
so 
we do not discuss the 
fully discrete case in this paper. 

\subsection{The Toda lattice in 
Flaschka--Manakov coordinates}

We consider the Toda lattice written 
in 
Flaschka--Manakov coordinates~\cite{Flaschka1,Flaschka2,Manakov74}: 
\begin{subequations}
\label{Toda-F}
\begin{align}
& u_{n,t} = u_n (v_{n} -v_{n-1}), 
\\
& v_{n,t} = u_{n+1} -u_{n}.
\label{Toda-F2}
\end{align}
\end{subequations}
The parametrization 
\mbox{$u_n = \mathrm{e}^{x_{n}-x_{n-1}}$}, 
\mbox{$\, v_n = x_{n,t}$} 
enables 
the system (\ref{Toda-F}) 
to be rewritten as the 
Newtonian equations of motion 
for the Toda lattice, 
\[
x_{n,tt} = \mathrm{e}^{x_{n+1}-x_{n}} - \mathrm{e}^{x_{n}-x_{n-1}}. 
\]
The 
Lax representation for the Toda lattice (\ref{Toda-F})~\cite{Suris03} 
can be written in the 
scalar (or ``big matrix'') form, 
\begin{subequations}
\label{Toda-LM}
\begin{align}
& \Psi_{n+1} + u_n \Psi_{n-1} =(\zeta + v_n ) \Psi_{n}, 
\label{Toda-L}
\\[0mm]
& \Psi_{n,t} = u_n \Psi_{n-1}.
\label{Toda-M}
\end{align}
\end{subequations}
Indeed, 
the 
commutativity of the spatial shift and the time derivative 
for $\Psi_n$ 
results in 
(\ref{Toda-F}). 
We introduce a
new variable $q_n$ as
\mbox{$\Psi_{n+1}/\Psi_{n}=: -q_n$}. 
Then, 
the Lax representation (\ref{Toda-LM})
implies the pair of relations 
\begin{subequations}
\begin{align}
& u_n = -(\zeta + v_n ) q_{n-1} - q_{n} q_{n-1}, 
\label{mT1}
\\
& q_{n,t} =-u_{n+1} -q_n \left[ (\zeta + v_{n} ) +q_{n} \right]. 
\label{mT2}
\end{align}
\end{subequations}
%
The first relation (\ref{mT1}) defines 
the Miura map 
\mbox{$(q_n,v_n) \mapsto (u_n,v_n)$}. 
Using (\ref{mT1}), 
we can eliminate $u_{n+1}$ and $u_n$ 
in 
(\ref{mT2}) and 
(\ref{Toda-F2}) to 
obtain a closed differential-difference 
system for 
\mbox{$(q_n,v_n)$}, 
\begin{subequations}
\label{mToda}
\begin{align}
& q_{n,t} = q_{n} (q_{n+1} + v_{n+1} - q_{n} - v_{n}),
\\
& v_{n,t} = -q_{n} ( q_{n+1} + v_{n+1} + \zeta ) 
	+q_{n-1} (q_{n} + v_n + \zeta ).
\end{align}
\end{subequations}
Note that the 
parameter $\zeta$
is 
no longer essential 
if we consider (\ref{mToda})
separately 
as an isolated system. Indeed, it can be set equal to zero 
by shifting $v_n$ as \mbox{$v_n + \zeta =: v_n'$}.
In addition, it is easy to identify (\ref{mToda}) 
with the Volterra lattice 
if we use the 
new 
pair of variables \mbox{$(-q_n, q_n+v_n+\zeta)$}.

The two-component 
system (\ref{Toda-F}) 
allows a one-parameter generalization 
corresponding to the relativistic deformation of 
the Toda lattice (see Chapter 6 
of~\cite{Suris03}). 
Thus, it would be interesting to 
obtain 
the 
one-parameter generalization of 
(\ref{mToda}) 
along the same lines 
as 
above 
and discuss its relationship with the 
relativistic Volterra lattice~\cite{Suris03}.

\subsection{The Belov--Chaltikian lattice}

The Belov--Chaltikian lattice~\cite{BC1} 
is described by the equations of motion
\begin{subequations}
\label{BC}
\begin{align}
& u_{n,t} = u_n (u_{n+1}-u_{n-1}) -(w_{n+1} -w_{n}), 
\label{BC-u}
\\
& w_{n,t} = w_n (u_{n+1} -u_{n-2}).
\label{BC-v}
\end{align}
\end{subequations}
The Lax representation for the 
Belov--Chaltikian lattice (\ref{BC})
is given by the 
pair of linear equations (cf.~\cite{HI97}), 
\begin{subequations}
\label{BC-LM}
\begin{align}
& \Psi_{n+1} -u_n \Psi_{n} + w_n \Psi_{n-1} = \zeta \Psi_{n+2}, 
\label{BC-L}
\\[0mm]
& \Psi_{n,t} = \Psi_{n+1} + u_{n-1} \Psi_{n}.
\label{BC-M}
\end{align}
\end{subequations}
If we introduce 
a new variable 
$v_n$ 
as
\mbox{$v_n :=\Psi_{n+1}/\Psi_{n}$},
the Lax representation (\ref{BC-LM})
implies the pair of relations
%
\begin{subequations}
\begin{align}
& w_n = ( \zeta v_{n+1} v_n - v_n +u_n ) v_{n-1}, 
\label{mBC1}
\\
& v_{n,t} = v_n (v_{n+1}+u_n) - v_n (v_n +u_{n-1}). 
\label{mBC2}
\end{align}
\end{subequations}
The first relation (\ref{mBC1}) defines 
the Miura map 
\mbox{$(u_n,v_n) \mapsto (u_n,w_n)$}. 
Using 
(\ref{mBC1}), 
we can eliminate $w_{n+1}$ and $w_n$
in 
(\ref{BC-u}); 
combining it with (\ref{mBC2}), we 
arrive at a closed differential-difference 
system for \mbox{$(u_n,v_n)$}, 
\begin{subequations}
\begin{align}
 u_{n,t} &= u_n (u_{n+1}-u_{n-1}) - u_{n+1}v_n +u_n v_{n-1}
\nonumber \\ 
& \hphantom{=} \;\,
	+v_{n} (v_{n+1}-v_{n-1} -\zeta v_{n+2} v_{n+1} 
	+\zeta v_{n+1} v_{n-1} ), 
\label{}
\\[1mm]
 v_{n,t} &= v_n (u_n-u_{n-1} +v_{n+1} -v_{n}).
\label{}
\end{align}
\end{subequations}
%
In the special case \mbox{$u_n =
0$}, 
(\ref{mBC1}) 
coincides with 
the Miura map 
proposed in~\cite{IW97} (also see~\cite{Kuper85}) 
for the Bogoyavlensky lattice, 
although 
the 
reduction 
\mbox{$u_n = 0$} 
is not consistent with 
this particular 
flow. 

\subsection{The relativistic Toda lattice}

The 
relativistic Toda lattice 
introduced by Ruijsenaars~\cite{Rui90},
\begin{equation}
x_{n,tt} = x_{n+1,t}x_{n,t} 
 \frac{g^2 \mathrm{e}^{x_{n+1}-x_{n}}}{1+g^2\mathrm{e}^{x_{n+1}-x_{n}} }
 - x_{n,t}x_{n-1,t}
  \frac{g^2 \mathrm{e}^{x_{n}-x_{n-1}}}{1+g^2 \mathrm{e}^{x_{n}-x_{n-1}}},
\label{rT-N}
\end{equation}
can be naturally written in the two-component form 
\begin{subequations}
\label{rT}
\begin{align}
& x_{n,t}=\mathrm{e}^{p_n} ( 1+g^2\mathrm{e}^{x_{n+1}-x_{n}} ),
\label{rT1}
\\
& p_{n,t}= g^2\mathrm{e}^{x_{n+1}-x_{n}+p_n}
	- g^2\mathrm{e}^{x_{n}-x_{n-1}+p_{n-1}}.
\label{rT2}
\end{align}
\end{subequations}
%
The \mbox{$2 \times 2$} Lax 
representation for (\ref{rT}) is given by~\cite{Suris9703}
\begin{subequations}
\label{rT-Lax}
\begin{align}
& 
\left[
\begin{array}{c}
 \Psi_{1,n+1} \\
 \Psi_{2,n+1} \\
\end{array}
\right]
=\left[
\begin{array}{cc}
 \zeta \mathrm{e}^{p_n} -\zeta^{-1} & \mathrm{e}^{x_n} \\
 -g^2 \mathrm{e}^{-x_n +p_n} & 0 \\
\end{array}
\right]
\left[
\begin{array}{c}
 \Psi_{1,n} \\
 \Psi_{2,n} \\
\end{array}
\right], 
\label{rT-Ln}
\\[1mm]
& \left[
\begin{array}{c}
 \Psi_{1,n} \\
 \Psi_{2,n} \\
\end{array}
\right]_t
= 
\left[
\begin{array}{cc}
 \zeta^{-2} + g^2 \mathrm{e}^{x_n -x_{n-1} +p_{n-1}} & 
 -\zeta^{-1} \mathrm{e}^{x_n} \\
 \zeta^{-1} g^2 \mathrm{e}^{-x_{n-1}+p_{n-1}} & 0 \\
\end{array}
\right]
\left[
\begin{array}{c}
 \Psi_{1,n} \\
 \Psi_{2,n} \\
\end{array}
\right].
\label{rT-Mn}
\end{align}
\end{subequations}
%
We 
introduce 
a new variable $y_n$ 
in the exponential form 
\mbox{$\mathrm{e}^{y_n} := \Psi_{1,n}/ \Psi_{2,n}$} 
so that a symmetric structure in 
the new pair 
of variables 
$x_n$ and $y_n$ 
can be uncovered. 
The Lax representation (\ref{rT-Lax}) 
enables 
$p_n$ and 
$y_{n,t}$ to be 
expressed 
as
\begin{subequations}
\label{rT-R}
\begin{align}
&  \mathrm{e}^{p_n} = \frac{\zeta^{-1}-\mathrm{e}^{x_n -y_{n}}}
	{\zeta + g^2 \mathrm{e}^{-x_n +y_{n+1}}},
\label{rT-R1}
\\[1mm]
&
y_{n,t} =
 \zeta^{-2} + g^2 \mathrm{e}^{x_n -x_{n-1} +p_{n-1}} 
 -\zeta^{-1} \mathrm{e}^{x_n-y_n} 
 -\zeta^{-1} g^2 \mathrm{e}^{-x_{n-1}+p_{n-1}+y_n}.
\label{rT-R2}
\end{align}
\end{subequations}
%
Thus, (\ref{rT-R1}) defines 
the Miura map \mbox{$(x_n,y_n) \mapsto (x_n,p_n)$}. 
Using (\ref{rT-R1}), we can eliminate $p_{n}$ 
in
(\ref{rT1}) and $p_{n-1}$ in 
(\ref{rT-R2}) to obtain a closed differential-difference
system for \mbox{$(x_n,y_n)$},
\begin{subequations}
\label{mrT}
\begin{align}
& x_{n,t}= \frac{(\zeta^{-1}-\mathrm{e}^{x_n -y_{n}})
( 1+g^2\mathrm{e}^{x_{n+1}-x_{n}} )}{\zeta + g^2 \mathrm{e}^{y_{n+1}-x_n}},
\label{mrT1}
\\[2mm]
& y_{n,t}= \frac{(\zeta^{-1}-\mathrm{e}^{x_n -y_{n}})
( 1+g^2\mathrm{e}^{y_{n}-y_{n-1}})}{\zeta + g^2 \mathrm{e}^{y_n -x_{n-1}}}.
\label{mrT2}
\end{align}
\end{subequations}
Actually, 
(\ref{mrT}) 
gives 
an auto-B\"acklund transformation 
between the two solutions $x_n$ and $y_n$ 
of the relativistic Toda lattice (\ref{rT-N}).

\subsection{The Ablowitz--Ladik lattice}
\label{secAL}

In this subsection, 
we 
consider 
the 
Ablowitz--Ladik lattice~\cite{AL1}, 
which 
appears 
to be 
the 
most instructive 
example 
in the discrete-space case. 
The equations of motion for
the (nonreduced form of the) 
Ablowitz--Ladik lattice 
are
\begin{subequations}
\label{sd-AL}
\begin{align}
& Q_{n,t}- a Q_{n+1} + b Q_{n-1} + (a-b)Q_n 
	+ a Q_{n+1} R_n Q_n - b Q_n R_n Q_{n-1} =O, 
\label{sd-AL1}
\\
& R_{n,t}- b R_{n+1} + a R_{n-1} + (b-a)R_n 
	+ b R_{n+1} Q_n R_n - a R_n Q_n R_{n-1} =O, 
\label{sd-AL2}
\end{align}
\end{subequations}
%
and 
the Lax representation 
is given by~\cite{AL1} 
\begin{subequations}
\label{AL-Lax}
\begin{align}
& 
\left[
\begin{array}{c}
 \Psi_{1,n+1} \\
 \Psi_{2,n+1} \\
\end{array}
\right]
=\left[
\begin{array}{cc}
 \zeta I_1 & Q_n \\
 R_n & \frac{1}{\zeta}I_2 \\
\end{array}
\right]
\left[
\begin{array}{c}
 \Psi_{1,n} \\
 \Psi_{2,n} \\
\end{array}
\right], 
\label{AL-Ln}
\\[1mm]
& \left[
\begin{array}{c}
 \Psi_{1,n} \\
 \Psi_{2,n} \\
\end{array}
\right]_t
= 
\left[
\begin{array}{cc}
 (\zeta^2 -1)a I_1 -aQ_n R_{n-1} & 
	\zeta a Q_n + \frac{b}{\zeta} Q_{n-1}\\
 \zeta a R_{n-1} + \frac{b}{\zeta} R_{n} & 
 \bigl( \frac{1}{\zeta^2} -1 \bigr) b I_2 -b R_n Q_{n-1} \\
\end{array}
\right]
\left[
\begin{array}{c}
 \Psi_{1,n} \\
 \Psi_{2,n} \\
\end{array}
\right].
\label{AL-Mn}
\end{align}
\end{subequations}
Here, $a$ and $b$ are free parameters. 
Note that
the Ablowitz--Ladik lattice (\ref{sd-AL}) is integrable for matrix-valued
dependent variables~\cite{GI82} 
(also see~\cite{Tsuchi02,
DM2010} 
and references therein);
in the
general case,
$Q_n$ and $R_n$ are
\mbox{$l_1 \times l_2$}
and \mbox{$l_2 \times l_1$} matrices, respectively.
We consider an \mbox{$(l_1+l_2) \times l_1$}
matrix-valued solution
to the pair of linear equations (\ref{AL-Lax})
such that $\Psi_{1,n}$ is an \mbox{$l_1 \times l_1$}
invertible
matrix.
Then,
in terms of
the \mbox{$l_2 \times l_1$} matrix 
\mbox{$P_n := \Psi_{2,n} \Psi_{1,n}^{-1}$},
(\ref{AL-Lax}) can be
rewritten as a pair of discrete and continuous 
matrix Riccati equations for $P_n$, 
\begin{subequations}
\label{AL-R}
\begin{align}
& R_n = \zeta P_{n+1} - \frac{1}{\zeta} P_n + P_{n+1}Q_n P_n,
\label{AL-R1}
\\[1mm]
& 
P_{n,t} =
 \zeta a R_{n-1} + \frac{b}{\zeta} R_{n} 
-(\zeta^2 -1)a P_n + \left( \frac{1}{\zeta^2} -1 \right) b P_n 
\nonumber \\[-1mm]
& \hphantom{P_{n,t} =}
-b R_n Q_{n-1} P_n + a P_n Q_n R_{n-1} 
- \frac{b}{\zeta} P_n Q_{n-1} P_n - \zeta a P_n Q_n P_n.
\label{AL-R2}
\end{align}
\end{subequations}
The first relation (\ref{AL-R1}) defines
the Miura map \mbox{$(Q_n,P_n) \mapsto (Q_n,R_n)$}. 
Using (\ref{AL-R1}), 
we can 
eliminate $R_n$ and $R_{n-1}$ in 
(\ref{sd-AL1}) and (\ref{AL-R2}) 
to obtain a closed differential-difference 
system for \mbox{$(Q_n,P_n)$} 
as (see (3.16) in~\cite{Tsuchi02})
\begin{subequations}
\label{sdGI}
\begin{align}
& Q_{n,t}- a Q_{n+1} + b Q_{n-1} + (a-b)Q_n 
	+ a Q_{n+1} 
\left( \zeta P_{n+1} - \frac{1}{\zeta} P_n \right) Q_n 
\nonumber \\
&
- b Q_n \left( \zeta P_{n+1} - \frac{1}{\zeta} P_n \right) Q_{n-1}
+ a Q_{n+1} P_{n+1}Q_n P_n Q_n 
- b Q_n  P_{n+1}Q_n P_n Q_{n-1}
=O, 
\label{}
\\[1mm]
& 
P_{n,t} -b P_{n+1} + a P_{n-1} +(b-a)P_n -b P_{n+1}\left( 
\frac{1}{\zeta} Q_n - \zeta Q_{n-1} \right) P_n
\nonumber \\
&
+a P_{n}\left( \frac{1}{\zeta} Q_n - \zeta Q_{n-1} \right) P_{n-1}
+ b P_{n+1} Q_n  P_{n}Q_{n-1} P_n - a P_n Q_{n} P_{n}Q_{n-1} P_{n-1}
=O.
\label{}
\end{align}
\end{subequations}
%
When \mbox{$a=-b=\mathrm{i}$}, 
(\ref{sdGI}) gives an integrable 
space discretization of the 
Gerdjikov--Ivanov system (\ref{mGI}).

Note that the parameter $\zeta$ in the modified system (\ref{sdGI}) is 
nonessential 
as in the continuous case. 
Indeed, for the elementary cases
\mbox{$a=0$} or \mbox{$b=0$}, 
a transformation of the form 
\[
Q_n= \zeta^{2n} {\mathrm e}^{ct} Q_n', \hspace{5mm}
P_n= \zeta^{-2n+1} {\mathrm e}^{-ct} P_n', 
\]
where $c$ is a suitably chosen parameter, 
and 
a rescaling of $t$
can be used 
to 
set $\zeta$ as $1$. 
For the more general case of \mbox{$ab \neq 0$}, 
we 
consider 
the redefinition 
of the parameters \mbox{$\zeta^2 a =a'$}, 
\mbox{$\zeta^{-2} b =b'$}, 
instead of rescaling $t$.

\section{\mbox{$(2+1)$}-dimensional PDEs}
\label{2D-PDE}

In this section, we 
illustrate 
how to apply our method in \mbox{$2+1$} 
dimensions. 
As 
instructive
examples, 
two 
distinct 
generalizations of 
the NLS system 
to \mbox{$2+1$} dimensions 
are 
considered. 

\subsection{\mbox{$(2+1)$}-dimensional NLS: 
Calogero--Degasperis system}
\label{sec-Zak}

In 
their 
pioneering
paper~\cite{Calo76}, 
Calogero and Degasperis proposed 
a large class of multidimensional 
PDEs 
that can be associated with the same spectral problem as 
the \mbox{$(1+1)$}-dimensional NLS system. 
A 
representative example of
the
class 
is 
a \mbox{$(2+1)$}-dimensional 
generalization of the 
NLS system (\ref{mNLS}), 
\begin{subequations}
\label{Z-sys}
\begin{align}
& \mathrm{i} Q_t + Q_{xy} - f Q -Qg= O,
\label{Za1}
\\
& \mathrm{i} R_t - R_{xy} + g R + R f = O,
\label{Za2}
\\
& f_x = (QR)_y, 
\label{Z-f}
\\
& g_x = (RQ)_y.
\label{Z-g}
\end{align}
\end{subequations}
Here, 
we consider 
the general case 
where the dependent variables 
are matrix-valued (cf.~\cite{Stra92}). 
Using (\ref{Z-f}) and (\ref{Z-g}), 
the auxiliary fields $f$ and $g$ 
can be formally written 
in the nonlocal forms
\mbox{$f=\partial_x^{-1}\partial_y (QR)$} 
and \mbox{$g=\partial_x^{-1}\partial_y (RQ)$}, 
which can be substituted 
back 
into (\ref{Za1}) and (\ref{Za2}). 
%
Note that by setting 
\mbox{$t = t_{n+1}$} and \mbox{$y = t_{n}$}~\cite{Gordoa99}, 
(\ref{Z-sys}) 
can be identified with 
the recursion relation 
for 
the 
\mbox{$(1+1)$}-dimensional NLS hierarchy~\cite{AKNS74}. 
Conversely, 
any \mbox{$(1+1)$}-dimensional integrable
system 
having 
a proper recursion operator 
can be generalized to \mbox{$2+1$} dimensions in this way. 

In the literature, 
the Calogero--Degasperis system (\ref{Z-sys}) is 
often 
referred to as the Zakharov system~\cite{Zakh}. 
The 
Lax representation for this system 
is given by~\cite{Zakh,Bogo91,Str92} (cf.\ (\ref{NLS-UV}))
\begin{subequations}
\label{Za-UV}
\begin{align}
& \left[
\begin{array}{c}
 \Psi_1  \\
 \Psi_2 \\
\end{array}
\right]_x
= \left[
\begin{array}{cc}
-\mathrm{i}\zeta I_1 & Q \\
 R & \mathrm{i}\zeta I_2\\
\end{array}
\right]
\left[
\begin{array}{c}
 \Psi_1  \\
 \Psi_2 \\
\end{array}
\right],
\label{Za-U}
\\[1mm]
& \left[
\begin{array}{c}
 \Psi_1  \\
 \Psi_2 \\
\end{array}
\right]_t
= 2 \zeta \left[
\begin{array}{c}
 \Psi_1  \\
 \Psi_2 \\
\end{array}
\right]_y 
+ 
\left[
\begin{array}{cc}
-\mathrm{i} f & \mathrm{i} Q_y \\
 -\mathrm{i} R_y & 
\mathrm{i} g \\
\end{array}
\right]
\left[
\begin{array}{c}
 \Psi_1  \\
 \Psi_2 \\
\end{array}
\right].
\label{Za-V}
\end{align}
\end{subequations}
%
Here, 
the 
spectral parameter $\zeta$, 
which is independent of $x$, 
has to satisfy 
the 
non-isospectral 
condition \mbox{$\zeta_t = 2\zeta \zeta_y$}~\cite{Bog90}. 
The Lax representation (\ref{Za-UV}) is 
valid 
for the simpler 
isospectral case 
\mbox{$\zeta_t = \zeta_y=0$}.
However, 
the 
non-isospectral nature 
of 
the 
parameter 
$\zeta$ 
turns out to be 
explicit 
and essential 
in applying the 
inverse 
scattering 
method~\cite{Calo76}. 

In terms of the 
new variable \mbox{$P := \Psi_2 \Psi_1^{-1}$},
the Lax representation (\ref{Za-UV}) can be
reformulated 
as a pair of matrix Riccati equations, 
\begin{subequations}
\label{Z-R}
\begin{align}
& P_x = R+ 2 \mathrm{i} \zeta P -PQP,
\label{Z-R1}
\\
& P_t - 2 \zeta P_y =- \mathrm{i}R_y + \mathrm{i}g P 
        + \mathrm{i}Pf - \mathrm{i}PQ_y P.
\label{Z-R2}
\end{align}
\end{subequations}
Thus, using (\ref{Z-R1}), 
which 
actually 
defines the Miura map
\mbox{$(Q,P) \mapsto (Q,R)$}, 
we can eliminate $R$ from 
(\ref{Z-f}), (\ref{Z-g}), and (\ref{Z-R2}) to obtain 
a closed 
system for 
\mbox{$(Q,P)$} 
with the auxiliary fields $f$ and $g$. 
To 
express 
this system in a more symmetric fashion, 
we 
redefine the auxiliary fields as 
\[
f=: -\frac{\mathrm{i}}{2} \zeta_y I_1 + \frac{1}{2} (QP)_y + u, 
\hspace{5mm}
g=: -\frac{\mathrm{i}}{2} \zeta_y I_2 + \frac{1}{2} (PQ)_y + v. 
\]
%
Then, 
we 
obtain
a \mbox{$(2+1)$}-dimensional generalization 
of the Gerdjikov--Ivanov system (\ref{mGI}) 
in the 
form 
\begin{subequations}
\label{mZ-sys}
\begin{align}
& \mathrm{i} Q_t + \mathrm{i} \zeta_y Q 
+ Q_{xy} - \left\{ u +\frac{1}{2}(QP)_y\right\}Q 
-Q\left\{ v +\frac{1}{2}(PQ)_y\right\}= O,
\label{2GI-Q}
\\
& \mathrm{i} P_t + \mathrm{i} \zeta_y P
- P_{xy} + \left\{ v -\frac{1}{2}(PQ)_y\right\} P 
+ P \left\{ u -\frac{1}{2}(QP)_y\right\} = O,
\label{2GI-P}
\\
& u_x =\left[ -2\mathrm{i}\zeta QP +\frac{1}{2}(QP_x-Q_xP)+(QP)^2 \right]_y, 
\label{}
\\
& v_x =\left[ -2\mathrm{i}\zeta PQ +\frac{1}{2}(P_xQ-PQ_x)+(PQ)^2 \right]_y,
\\
& \zeta_t = 2\zeta \zeta_y, \hspace{5mm} \zeta_x=0. 
\label{}
\end{align}
\end{subequations}
It is 
now 
clear how 
to 
impose suitable reductions,  
such as the complex 
conjugacy reduction, 
on the dependent variables. 
In the 
case of \mbox{$\zeta_t = \zeta_y=0$}, 
(\ref{2GI-Q}) and (\ref{2GI-P}) imply 
the conservation law
\[
\mathrm{i}
\frac{\partial}{\partial t} \mathrm{tr}\hspace{1pt}(QP)
 + \frac{\partial}{\partial x} \mathrm{tr}\left[ 
\frac{1}{2}(Q_{y}P-QP_y )\right]
 + \frac{\partial}{\partial y}
        \mathrm{tr}\left[ \frac{1}{2}(Q_{x}P-QP_x) -(QP)^2 \right]
= 0.
\]
%
Using the same 
technique as 
in the \mbox{$(1+1)$}-dimensional case (see 
(\ref{P-series})), 
we can construct an infinite set of conservation laws for 
the Calogero--Degasperis system (\ref{Z-sys})
(cf.~\cite{Bull94}). 
Thus, the conservation laws for (\ref{mZ-sys}) 
in the general 
case of nonconstant 
$\zeta$ can 
be obtained 
with the aid of 
(\ref{Z-R1}). 

\subsection{\mbox{$(2+1)$}-dimensional NLS: Davey--Stewartson system}
\label{sec-DS}

The other \mbox{$(2+1)$}-dimensional generalization of the
NLS system (\ref{mNLS}) to be considered is 
(the integrable 
case 
of) the Davey--Stewartson system~\cite{DS74}, 
also 
referred to 
as the 
Benney--Roskes system~\cite{Benney69}; 
it can be written in the form 
\begin{subequations}
\label{DS-sys}
\begin{align}
& \mathrm{i} Q_t + a Q_{xx} +bQ_{yy} - f Q -Qg= O,
\label{DS1}
\\
& \mathrm{i} R_t - a R_{xx} - bR_{yy} + g R + R f = O,
\label{DS2}
\\
& f_x = 2b(QR)_y, 
\label{DS-f}
\\
& g_y = 2a(RQ)_x.
\label{DS-g}
\end{align}
\end{subequations}
Here, $a$ and $b$ are free parameters, 
and 
$f$ and $g$ are auxiliary fields. 
The simplest case 
of 
\mbox{$b=0$} 
(or \mbox{$a=0$}) 
provides a \mbox{$(2+1)$}-dimensional generalization 
of the Yajima--Oikawa system (\ref{YOsys})~\cite{Zakh,Mel83}. 
Note that 
the Davey--Stewartson system 
(\ref{DS-sys}) 
is 
integrable 
for matrix-valued dependent variables~\cite{Kono92,Mel83,Fordy87,March}. 
Its Lax representation 
is given by~\cite{Hab75,Anker,Ab78,Zakh}
%
\begin{subequations}
\label{DS-UV}
\begin{align}
& \left[
\begin{array}{c}
\partial_x \Psi_{1}  \\
\partial_y \Psi_{2} \\
\end{array}
\right]
= \left[
\begin{array}{cc}
 O & Q \\
 R & O \\
\end{array}
\right]
\left[
\begin{array}{c}
 \Psi_1  \\
 \Psi_2 \\
\end{array}
\right],
\label{DS-U}
\\[1mm]
& \mathrm{i}
\left[
\begin{array}{c}
 \Psi_1  \\
 \Psi_2 \\
\end{array}
\right]_t
= 
\left[
\begin{array}{cc}
-b \partial_y^2 + f & a Q \partial_x -a Q_x \\
 -b R \partial_y + b R_y &  a \partial_x^2 - g \\
\end{array}
\right]
\left[
\begin{array}{c}
 \Psi_1  \\
 \Psi_2 \\
\end{array}
\right].
\label{DS-V}
\end{align}
\end{subequations}
%
In contrast with (\ref{Za-U}), 
the spatial part 
of the 
Lax representation, (\ref{DS-U}), 
is a two-dimensional 
problem 
without a spectral parameter. 

Let us introduce a
new variable \mbox{$P := \Psi_2 \Psi_1^{-1}$}. 
Owing to
(\ref{DS-U}), we can introduce the auxiliary field $w$ as
\begin{equation}
\Psi_{1,x} = QP \Psi_1,\hspace{5mm}
\Psi_{1,y} =  w \Psi_1.
\label{QP-w}
\end{equation}
Note that \mbox{$(\Psi_1^{-1})_x = -\Psi_1^{-1}QP$}. 
The compatibility condition of
these linear
PDEs for $\Psi_1$ implies 
the relation
\begin{equation}
w_x - (QP)_y + \left[ w, QP \right] =O.
\label{w-QP}
\end{equation}
From (\ref{DS-U}) and (\ref{QP-w}), we obtain 
\begin{equation}
R= P_y + Pw, 
\label{R-Pw}
\end{equation}
which
defines the Miura map \mbox{$(Q,P) \mapsto (Q,R)$}. 
Thus, equations (\ref{DS-f}) and (\ref{DS-g}) for the 
auxiliary fields $f$ and $g$ can be rewritten as 
\begin{align}
f_x &= 2b\left( Q P_y + QPw \right)_y,
\label{DS-f'} \\
g_y &= 2a \left( P_y Q + Pw Q \right)_x.
\label{DS-g'}
\end{align}
Noting the identity 
\[
\Psi_{2,xx} \Psi_1^{-1} = (\Psi_2 \Psi_1^{-1})_{xx} 
 - 2 (\Psi_{2} \Psi_1^{-1})_x \Psi_{1} (\Psi_1^{-1})_x 
 + 2 \Psi_{2} (\Psi_1^{-1})_x \Psi_{1} (\Psi_1^{-1})_x
 - \Psi_{2} (\Psi_1^{-1})_{xx}
\]
and using (\ref{DS-V}), we can compute the time derivative of 
\mbox{$P (= \Psi_2 \Psi_1^{-1})$}
with the aid of (\ref{QP-w}) and (\ref{R-Pw}) as 
\begin{align}
\mathrm{i} P_t &= a \Psi_{2,xx} \Psi_1^{-1} -g \Psi_2 \Psi_1^{-1}
	-b R \Psi_{1,y} \Psi_1^{-1} + bR_y 
	+b\Psi_2 \Psi_1^{-1}\Psi_{1,yy} \Psi_1^{-1} 
\nonumber \\ & 
\hphantom{=} \;\,
	-\Psi_2 \Psi_1^{-1} f 
	-a\Psi_2 \Psi_1^{-1} Q \Psi_{2,x} \Psi_1^{-1} 
	+a\Psi_2 \Psi_1^{-1}Q_x \Psi_2 \Psi_1^{-1}
\nonumber \\[1mm] & = 
a P_{xx} + 2a P_x QP + 2a PQPQP -a PQPQP + aP(QP)_x 
\nonumber \\ & 
\hphantom{=} \;\,
-gP -b P_y w 
-b Pw^2 +b P_{yy} +b(Pw)_y +bPw_y  +bPw^2 
\nonumber \\ & 
\hphantom{=} \;\,
-Pf -aPQP_x -aPQPQP +a PQ_x P
\nonumber \\[1mm] & = a P_{xx} +bP_{yy} -Pf -gP +2bPw_y +2a (PQ)_x P. 
\label{mDS-P}
\end{align}
Thus, (\ref{DS1}), (\ref{mDS-P}), (\ref{DS-f'}), and (\ref{DS-g'}) 
together with (\ref{w-QP}) comprise the modified system. 
To
rewrite
it
in a more symmetric fashion,
we
redefine the auxiliary fields 
as
\[
f=: bw_y - H,
\hspace{5mm}
g=: a (PQ)_x -F .
\]
Then,
we
obtain
a
\mbox{$(2+1)$}-dimensional generalization
of the Gerdjikov--Ivanov system (\ref{mGI})
in the
form
\begin{subequations}
\label{mDS-sys}
\begin{align}
& \mathrm{i} Q_t + a Q_{xx} +bQ_{yy} + Q \left[ F -a (PQ)_x \right]
	+ \left( H-bw_y \right) Q = O,
\\
& \mathrm{i} P_t - a P_{xx} -bP_{yy}- \left[ F +a (PQ)_x \right] P
	-P \left( H +bw_y \right) =O,
\\
& F_y + a \left( P_y Q-PQ_y  + 2Pw Q \right)_x =O,
\\
& H_x + b\left( Q P_y -Q_y P + QPw +wQP \right)_y =O,
\\
& w_x - (QP)_y + \left[ w, QP \right] =O.
\end{align}
\end{subequations}
For the 
scalar 
(and thus commutative) case, 
this system 
and 
the Miura map 
to the Davey--Stewartson system
were  
derived in~\cite{MY97} using a different approach. 

%
%
\section{Concluding remarks}

In this paper, we have
developed an
effective method of
identifying 
new integrable systems
that 
can be 
mapped 
to a given 
integrable 
system 
by 
Miura
transformations. 
The method 
is applicable in a systematic manner
%
as long as 
%
the spatial part of the 
Lax representation for
the 
given 
system is ultralocal in the 
dependent 
variables. 
In fact, most of the known integrable systems 
satisfy this requirement, 
and the wide applicability of the method 
is illustrated using numerous
examples. 
%
The cornerstone of our method is 
to overcome a stereotype of perceiving 
the original 
dependent variables 
and an eigenfunction of the associated 
linear problem as 
entirely different 
objects that 
cannot be swapped. 
That is, we 
combine 
a subset 
of the original 
dependent 
variables 
and 
the components of 
an 
eigenfunction of the 
linear problem 
to 
form 
a new 
set of 
dependent variables 
that satisfies
a 
closed system; 
this process 
elucidates the true nature of the Lax representation 
and 
explains its origin.
In the appendices, 
a variant of the method is 
used 
to derive and characterize
derivative NLS systems 
of the Chen--Lee--Liu type. 

%
%

To illustrate the method, 
we mainly 
considered 
the first nontrivial 
flows of 
integrable hierarchies 
as 
examples. 
Note, however, that the method can also be applied to 
the 
higher flows in each integrable hierarchy, 
as well as 
to the 
negative flows. 
Indeed, 
a
Miura map between two 
integrable systems 
actually gives a link between the original hierarchy 
and the modified hierarchy
rather than between a particular 
flow 
and its 
modification. 
Thus, 
an integrable system is 
not an isolated object 
either 
``longitudinally'' 
or ``transversely''; 
it not only arises as a member 
in 
an infinite hierarchy 
of commuting flows, but also appears 
with its partners
related by a chain of Miura maps.
That is, a given integrable system 
is either 
an original system 
having 
one or more 
modifications 
or a 
modification of a more 
basic 
system 
or both. 

Originally
invented to derive
new integrable systems 
from known ones,
our method also 
provides the most direct route to
solving the 
derived 
integrable 
systems. 
By construction, 
the 
set of 
dependent variables satisfying 
a derived modified system 
is written 
explicitly in terms of the original 
variables and 
an eigenfunction of the 
linear problem 
for the original system. 
Moreover, 
when the inverse scattering method 
is 
applied to the original nonlinear system, 
it 
intrinsically 
provides 
formulas for determining 
the eigenfunctions 
of the associated linear 
problem. 
Thus, if the original system can be 
solved by the inverse 
scattering method, 
then 
it is 
an easy and straightforward task 
to 
obtain 
a solution formula for the modified system using 
simple operations such as division and 
integration by parts. 
Incidentally, this kind of operation plays a critical role in integrating 
the sine-Gordon equation and its 
discretizations explicitly. 
Note, however, that 
the general solution of 
an \mbox{$l \times l$} 
matrix 
or $l$-th-order scalar 
linear problem 
is expressed as 
a
linear combination 
of $l$ 
fundamental solutions 
with $l$ arbitrary coefficients. 
Therefore, we need 
to impose suitable boundary conditions 
on the modified system 
to remove this arbitrariness and 
obtain 
its solution formula
uniquely. 
%
%
%

For a \mbox{$(1+1)$}-dimensional 
system, 
the existence of an arbitrary 
parameter, 
called the spectral parameter, 
in the Lax representation 
is a manifestation of 
its integrability. 
Consequently, 
the Miura map and the modified system 
obtained 
by 
our method 
naturally contain 
the spectral parameter 
as a 
free parameter. 
In general, the spectral parameter has its origin in 
a group of point transformations 
that leave the original unmodified hierarchy invariant. 
For example, the spectral problem (\ref{NLS-U}) is 
invariant under the one-parameter group of transformations: 
\mbox{$\Psi_1':= \Psi_1 \mathrm{e}^{\mathrm{i} k x} $}, 
\mbox{$\Psi_2':= \Psi_2 \mathrm{e}^{-\mathrm{i} k x} $}, 
\mbox{$Q':= Q\mathrm{e}^{2\mathrm{i} k x} $}, 
\mbox{$R':= R\mathrm{e}^{-2\mathrm{i} k x} $}, and 
\mbox{$\zeta' := \zeta - k $}. 
Thus, 
the associated 
NLS hierarchy is also invariant under the 
transformation of dependent variables, 
\mbox{$Q'= Q\mathrm{e}^{2\mathrm{i} k x} $} and 
\mbox{$R'= R\mathrm{e}^{-2\mathrm{i} k x} $}. 
Note that 
this transformation 
induces 
a linear change in 
the 
infinite set of 
time variables 
for 
the NLS hierarchy, 
and 
thus, 
each flow 
is not invariant. 
The invariance of 
individual
NLS flows can be restored by 
applying the inverse of 
this linear 
change 
in 
the time variables. 
In 
the simplest 
case of the first nontrivial 
flow, 
the NLS system (\ref{mNLS}), this 
amounts to 
its 
Galilean invariance~\cite{SY74}. 
This implies that 
the parameter $\zeta$ in the corresponding modified system (\ref{mGI}) 
can be fixed 
at any value 
by 
a point transformation, cf.~appendix~\ref{app1}\@. 
In general, 
the spectral parameter appearing 
in 
the Miura map is nonessential 
in the sense that it 
can be 
fixed at any generic value 
using
a 
point transformation. 
However, 
such a point transformation usually 
does not leave the modified integrable hierarchy invariant, 
so 
the spectral parameter 
can be viewed as a deformation parameter of the modified hierarchy. 

To avoid needless confusion, we did not proactively address the issues 
of certain 
types 
of nonstandard Lax representations; 
in fact, 
there exist 
cases wherein 
the components of the 
linear 
eigenfunction 
in the matrix 
Lax representation 
are not 
entirely independent. 
Thus, we have to identify 
an appropriate subset of the 
components to 
define 
the relevant Miura map correctly; 
this point was
briefly and partially 
touched upon in section~\ref{sect2}. 
There are two 
main types 
of such 
nonstandard 
Lax representations. 
In the first type,
linear ordinary differential/difference 
relations among 
components 
with respect to the spatial variable, 
such as 
\mbox{$\Psi_3 = \partial_x^2 \Psi_{1}$}, 
hold true. 
Typical 
examples of this type 
are 
high-order 
scalar Lax 
pairs 
reformulated 
in the matrix form. 
In the second type, we 
have 
nonlinear algebraic relations  
among 
components, 
such as \mbox{$(\Psi_2)^2 =\Psi_1 \Psi_3$}, 
so that 
not all 
components are 
algebraically 
independent. 
For example, the scalar case of the NLS system (\ref{mNLS}) 
allows
the standard Lax representation 
in the \mbox{$2 \times 2$} matrix form (see (\ref{NLS-UV})); 
however, it 
also 
implies a \mbox{$3 \times 3$} 
nonstandard Lax representation 
for the ``squared''
eigenfunction 
\mbox{$\left( (\Psi_1)^2, \Psi_1 \Psi_2, (\Psi_2)^2 \right)^T$}~\cite{Ab78} 
and 
a \mbox{$4 \times 4$} 
nonstandard 
Lax representation 
for the 
``cubed'' eigenfunction with components 
\mbox{$(\Psi_1)^{3-j} (\Psi_2)^{j}$}, \mbox{$j=0,1,2,3$} 
and so forth.
%
%
For a given Lax representation, 
there exists 
no 
algorithmic 
way to 
find 
all 
nontrivial 
relations among 
components 
of the 
eigenfunction. 
However, 
it is often possible 
to 
identify 
such 
relations 
by 
noting 
the 
internal 
symmetry
of the 
Lax 
pair. 
For 
example, 
if the Lax 
matrices 
are antisymmetric
up to a certain similarity transformation, 
with respect to the secondary diagonal, 
then a quadratic relation among the components results
(cf.~\cite{Adler08,Adler94}). 

%
%

As a final remark, we note that
our method can 
also 
be applied to 
classical many-body
problems such as 
the Calogero--Moser
models. 
In that case, 
the method 
pinpoints 
a nontrivial 
transformation 
from the 
modified system 
to 
a given original system. 
%
The 
transformation 
is 
no longer a Miura map in the usual sense 
of the word; 
rather, it is 
a coordinate transformation 
acting on the 
finite set of dynamical variables 
\mbox{$\{\vt{q},\vt{p}\}$}. 
The existence of such a transformation is highly nontrivial; 
it maps the modified system to the original system 
in an easy-to-follow manner,
but its inverse 
cannot be found without 
using a systematic approach. 
%
%

\section*{Acknowledgments} 
Folkert M\"uller-Hoissen, 
Stephen Anco, 
Kouichi Toda, 
and 
Tadashi Kobayashi 
are thanked for 
their useful comments 
and/or
discussions. 

\appendix
\section{Continuous Chen--Lee--Liu system}
\label{app1}
In this appendix, we show 
that 
the Lax representation for 
the NLS system 
can 
provide 
the solutions of 
a derivative NLS system, called the Chen--Lee--Liu system~\cite{CLL}. 
This result 
can be obtained 
by 
exploiting the transformation of dependent variables 
between 
the Gerdjikov--Ivanov system (\ref{mGI}) and 
the Chen--Lee--Liu system. 

Let $Q$ and $R$ satisfy the nonreduced 
NLS system (\ref{mNLS}). 
Let $\Psi_1$ and $\Psi_2$ be the 
first and second components of 
an eigenfunction 
of the associated linear problem
(\ref{NLS-UV}). 
Then, we have the following proposition.
\vspace{5mm}
\begin{proposition}
\label{propA.1}
{\em 
The new pair of variables,
\begin{equation}
\left\{
\begin{split}
& q
:= \left( \Psi_1 {\mathrm e}^{\mathrm{i} \zeta x 
	+2 \mathrm{i} \zeta^2 t} \right)^{-1} Q, 
\\
& r
:= \Psi_2 {\mathrm e}^{\mathrm{i} \zeta x 
	+2 \mathrm{i} \zeta^2 t}, 
\end{split}
\right.
\label{NLS-CLL}
\end{equation}
satisfies a closed system, i.e., the 
derivative NLS system
\begin{equation}
\left\{
\begin{split}
& \mathrm{i} q_t + q_{xx} 
 + 4\mathrm{i}\zeta qrq + 2 qrq_x = O,
\\
& \mathrm{i} r_t - r_{xx} 
 - 4\mathrm{i}\zeta rqr + 2 r_x qr = O. 
\end{split}
\right.
\label{mCLL}
\end{equation}
}
\end{proposition}
\vspace{5mm}
\noindent
{\bf Remark.}~~The exponential factor 
\mbox{${\mathrm e}^{\mathrm{i} \zeta x +2 \mathrm{i} \zeta^2 t}$}
is introduced in (\ref{NLS-CLL}) simply to remove nonessential 
linear terms from the resulting system, (\ref{mCLL}). 
The standard form of the Chen--Lee--Liu system 
corresponds to 
the 
case of 
\mbox{$\zeta=0$} 
(see~\cite{Linden1,Olver2,TW3,Ad,Dimakis} for the matrix
case). 
Note that 
the parameter $\zeta$ 
is nonessential 
and 
(\ref{mCLL}) 
can be reduced to this case 
by a point transformation. 
Indeed, if we change the variables in (\ref{NLS-UV}) as 
\[
\begin{array}{l}
\Psi_1' := \Psi_1 {\mathrm e}^{\mathrm{i}\zeta x +2\mathrm{i}\zeta^2 t}, 
\hspace{3mm}
\Psi_2' := \Psi_2 {\mathrm e}^{-\mathrm{i}\zeta x -2\mathrm{i}\zeta^2 t}, 
\hspace{3mm}
Q' := Q {\mathrm e}^{2\mathrm{i}\zeta x +4\mathrm{i}\zeta^2 t}, \hspace{3mm}
R' := R {\mathrm e}^{-2\mathrm{i}\zeta x -4\mathrm{i}\zeta^2 t}, 
\\[2mm]
\partial_{t'} := \partial_{t} -4\zeta \partial_x, 
\hspace{3mm}
\partial_{x'} := \partial_{x}
\end{array}
\]
and omit the prime,
we obtain the same Lax representation 
(\ref{NLS-UV}) 
with \mbox{$\zeta=0$}. 
This is a manifestation of the Galilean invariance
of the NLS system (\ref{mNLS})~\cite{SY74}. 
\vspace{5mm}
\\
\noindent
{\bf Proof.}~~For brevity, we 
use the 
quantities \mbox{$\Phi_j :=\Psi_j {\mathrm e}^{\mathrm{i} \zeta x 
	+2 \mathrm{i} \zeta^2 t} \; (j=1,2)$} 
instead of $\Psi_j$.
Using the Lax representation (\ref{NLS-UV}), 
we can 
express 
the time derivatives of $\Phi_j$ 
as
\begin{align}
\mathrm{i}
\Phi_{1,t} &=
QR \Phi_1 + 2\mathrm{i}\zeta Q \Phi_2 -Q_x \Phi_2 
\nonumber \\
&= - \left( Q \Phi_2 \right)_x + 4\mathrm{i}\zeta Q \Phi_2 + 2QR \Phi_1
\nonumber \\
&= - \Phi_{1,xx} + 4\mathrm{i}\zeta Q \Phi_2 + 2QR \Phi_1, 
\label{Phi1-t} 
\\[3mm]
\mathrm{i}
\Phi_{2,t}
&= 2\mathrm{i}\zeta R \Phi_1 + R_x \Phi_1 - 4\zeta^2 \Phi_2
	 -RQ\Phi_2
\nonumber \\
&=(R\Phi_1)_x + 2\mathrm{i}\zeta\Phi_{2,x} -2RQ \Phi_2
\nonumber \\
&= \Phi_{2,xx} -2(\Phi_{2,x} - 2\mathrm{i}\zeta\Phi_2) \Phi_1^{-1}Q \Phi_2. 
\label{Phi2-t}
\end{align}
Because \mbox{$\Phi_2=r$} and \mbox{$\Phi_1^{-1}Q=q$}, 
(\ref{Phi2-t}) 
can be identified with 
the second equation in (\ref{mCLL}). 
Combining (\ref{Phi1-t}) with (\ref{NLS1}) and 
noting the relation \mbox{$\Phi_{1,x}=Q\Phi_2$} 
(cf.~(\ref{NLS-U})), we can 
rewrite the time derivative of \mbox{$\Phi_1^{-1}Q$} as 
\begin{align}
\mathrm{i}
\left( \Phi_1^{-1}Q \right)_t &=
\Phi_1^{-1} \Phi_{1,xx} \Phi_1^{-1}Q 
-4\mathrm{i}\zeta \Phi_1^{-1}Q \Phi_2 \Phi_1^{-1}Q 
-\Phi_1^{-1} Q_{xx}
\nonumber \\
&= \left( \Phi_1^{-1} \Phi_{1,x} \Phi_1^{-1}\right)_x Q 
+2 \Phi_1^{-1} \Phi_{1,x} \Phi_1^{-1} \Phi_{1,x} \Phi_1^{-1}Q 
-\Phi_1^{-1} Q_{xx}
\nonumber \\ & \hphantom{=} \;\,
-4\mathrm{i}\zeta \Phi_1^{-1}Q \Phi_2 \Phi_1^{-1}Q 
\nonumber \\
&= -\left( \Phi_1^{-1} Q \right)_{xx} 
-2 \Phi_1^{-1} Q\Phi_2 \left( \Phi_1^{-1}Q \right)_x
-4\mathrm{i}\zeta \Phi_1^{-1}Q \Phi_2 \Phi_1^{-1}Q. 
\nonumber
\end{align}
This verifies the first equation in (\ref{mCLL}). 
\hfill $\Box$
\vspace{5mm}

The correspondence relation 
(\ref{NLS-CLL}) between the NLS system 
and the Chen--Lee--Liu system 
can be generalized
for 
the higher/negative
flows of 
the 
hierarchies. 
In addition, it is easy 
to extend 
this idea 
to the case of \mbox{$3 \times 3$} 
(or even higher)
matrix 
Lax representations, 
{\it e.g.}, (\ref{3W-UV})
for the three-wave interaction system (\ref{3W}). 

\section{Semi-discrete Chen--Lee--Liu system}
\label{app2}

In this appendix, we show 
that 
the Lax representation for the Ablowitz--Ladik lattice 
can provide 
the solutions of 
an integrable 
space discretization 
(semi-discretization, for short) of 
the Chen--Lee--Liu system. 
This result 
can 
be 
found 
by 
exploiting the transformation of dependent variables 
between 
the semi-discrete Gerdjikov--Ivanov system (\ref{sdGI}) and 
the semi-discrete Chen--Lee--Liu system. 

Let $Q_n$ and $R_n$ satisfy the nonreduced Ablowitz--Ladik lattice 
(\ref{sd-AL}). 
Let $\Psi_{1,n}$ and $\Psi_{2,n}$ be the 
first and second components of 
an eigenfunction 
of the associated linear problem
(\ref{AL-Lax}). 
Then, we can state 
the following 
discrete 
analog
of Proposition~\ref{propA.1}. 
\vspace{5mm}
\begin{proposition}
\label{prop.B.1}
{\em 
The new pair of variables, 
\begin{equation}
\left\{
\begin{split}
& q_n 
:= \left( \Psi_{1,n} \zeta^{-n} {\mathrm e}^{
-(1/\zeta^2 -1)b t} 
 \right)^{-1} Q_{n-1}, 
\\[1mm]
& r_n 
:= \Psi_{2,n} \zeta^{-n} {\mathrm e}^{-(1/\zeta^2 -1)b t}, 
\\
\end{split}
\right.
\label{AL-CLL}
\end{equation}
satisfies a closed system, i.e., the 
semi-discrete 
Chen--Lee--Liu system~{\rm{\cite{Tsuchi02}}}
\begin{equation}
\left\{
\begin{split}
& q_{n,t} -a \left( I - \frac{1}{\zeta}q_{n+1}r_n \right)^{-1} 
 \left( q_{n+1}- \zeta^2 q_n \right) 
 -b \left( I - \zeta q_n r_n \right) 
 \left( \frac{1}{\zeta^2}q_n - q_{n-1} \right) = O,
\\
& r_{n,t} -b \left( r_{n+1}-\frac{1}{\zeta^2} r_n \right) 
	\left( I -\zeta q_{n} r_n \right) -a (\zeta^2 r_{n}-r_{n-1})
 \left( I-\frac{1}{\zeta} q_{n} r_{n-1} \right)^{-1} = O. 
\end{split}
\right.
\label{dCLL}
\end{equation}
Moreover, multiplying 
\mbox{$\Psi_{j,n}$} by 
factor \mbox{$\mathrm{e}^{ct}$} 
in the original definition of $q_n$ and $r_n$ in \mbox{\rm{(\ref{AL-CLL})}}, 
one can add the terms 
\mbox{$ 
+c q_n$} and \mbox{$-c r_n$} 
to the left-hand side of 
equations \mbox{\rm (\ref{dCLL})}. 
The nonzero parameter $\zeta$ is nonessential, 
because it can be 
set as
$1$ by the transformation 
\mbox{$q_n =: q_n' \zeta^{2n-1}$}, \mbox{$r_n =: r_n' \zeta^{-2n}$}
and the 
redefinition 
of the parameters 
\mbox{$a\zeta^2 =: a'$}, 
\mbox{$b/\zeta^2 =: b'$}.
}
\end{proposition}
\vspace{5mm}
\noindent
{\bf Remark.}~~The 
index 
of the unit matrix $I$ 
to indicate its size 
is suppressed 
in (\ref{dCLL}) and below. 
For scalar $q_n$ and $r_n$, 
(\ref{dCLL}) 
was 
previously 
studied 
in~\cite{SY91,AY94,ASY00} 
(also see~\cite{DJM83}). 
Clearly, 
(\ref{dCLL}) with \mbox{$a=-b=\mathrm{i}$} is not 
an ideal 
semi-discretization of 
(\ref{mCLL}), 
because it does not allow
a 
local reduction 
to relate $q_n$ and $r_n$ by 
complex/Hermitian conjugation. 
However, (\ref{dCLL}) 
in the scalar case 
is an interesting 
system 
possessing 
an 
ultralocal Hamiltonian 
structure~\cite{SY91,AY94,ASY00}, which 
plays a role in discussions on
a tri-Hamiltonian structure of the Ablowitz--Ladik 
lattice and related systems 
(cf.\ the concluding remarks in~\cite{Tsuchi02}). 
\vspace{5mm}
\\
\noindent
{\bf Proof.}~~For brevity, we
use the
quantities \mbox{$\Phi_{j,n} :=
\Psi_{j,n} \zeta^{-n} {\mathrm e}^{
-(1/\zeta^2 -1)b t}$} that satisfy the 
following linear problem 
(cf.~(\ref{AL-Lax})): 
\begin{subequations}
\label{AL-Lax2}
\begin{align}
&
\left[
\begin{array}{c}
 \Phi_{1,n+1} \\
 \Phi_{2,n+1} \\
\end{array}
\right]
=\left[
\begin{array}{cc}
 I & \frac{1}{\zeta} Q_n \\
 \frac{1}{\zeta} R_n & \frac{1}{\zeta^2}I \\
\end{array}
\right]
\left[
\begin{array}{c}
 \Phi_{1,n} \\
 \Phi_{2,n} \\
\end{array}
\right],
\label{AL-Ln2}
\\[1mm]
& \left[
\begin{array}{c}
 \Phi_{1,n} \\
 \Phi_{2,n} \\
\end{array}
\right]_t
=
\left[
\begin{array}{cc}
 \left\{ \left( \zeta^2 -1 \right) a - \left( \frac{1}{\zeta^2} -1 \right) 
  b \right\} I -aQ_n R_{n-1} & \zeta a Q_n + \frac{b}{\zeta} Q_{n-1}\\
 \zeta a R_{n-1} + \frac{b}{\zeta} R_{n} & -b R_n Q_{n-1} \\
\end{array}
\right]
\left[
\begin{array}{c}
 \Phi_{1,n} \\
 \Phi_{2,n} \\
\end{array}
\right].
\label{AL-Mn2}
\end{align}
\end{subequations}
Using (\ref{AL-Lax2}) and (\ref{sd-AL1}), 
we can rewrite 
the time derivatives of 
\mbox{$\Phi_{1,n}^{-1}Q_{n-1}$} and $\Phi_{2,n}$ as 
\begin{subequations}
\label{AL-phi}
\begin{align}
&  
\left( \Phi_{1,n}^{-1} Q_{n-1}\right)_t
\nonumber \\
=& - \!\left[ \left(\zeta^2-1 \right)a - \left( \frac{1}{\zeta^2}-1  \right) b 
\right] \Phi_{1,n}^{-1} Q_{n-1} 
+ a\Phi_{1,n}^{-1} Q_n R_{n-1}Q_{n-1}
\nonumber \\ & 
- \zeta a \Phi_{1,n}^{-1} Q_n \Phi_{2,n}\Phi_{1,n}^{-1} Q_{n-1}
- \frac{b}{\zeta} \Phi_{1,n}^{-1} Q_{n-1}\Phi_{2,n}\Phi_{1,n}^{-1} Q_{n-1}
\nonumber \\ 
& 
+ \Phi_{1,n}^{-1} \left[ a Q_{n} - b Q_{n-2} - (a-b)Q_{n-1}
        - a Q_{n} R_{n-1} Q_{n-1} + b Q_{n-1} R_{n-1} Q_{n-2} \right]
\nonumber \\[1mm]
=& - \! \zeta^2 a \Phi_{1,n}^{-1} Q_{n-1} 
+ a \Phi_{1,n}^{-1} Q_n \left(I- 
	\zeta \Phi_{2,n}\Phi_{1,n}^{-1} Q_{n-1} \right)
\nonumber \\ & 
+ \frac{b}{\zeta^2}\Phi_{1,n}^{-1} Q_{n-1} 
\left( I -\zeta \Phi_{2,n}\Phi_{1,n}^{-1} Q_{n-1} \right)
- b  \Phi_{1,n}^{-1} Q_{n-2} + b  \Phi_{1,n}^{-1} Q_{n-1} R_{n-1} Q_{n-2} 
\nonumber \\
=& -\! \zeta^2 a \Phi_{1,n}^{-1} Q_{n-1}
+ a \left(I-\frac{1}{\zeta} \Phi_{1,n+1}^{-1}Q_n \Phi_{2,n} \right)^{-1} 
	\Phi_{1,n+1}^{-1} Q_n \left(I- 
	\zeta \Phi_{2,n}\Phi_{1,n}^{-1} Q_{n-1} \right)
\nonumber \\ & 
+ \frac{b}{\zeta^2}
\left( I -\zeta \Phi_{1,n}^{-1} Q_{n-1} \Phi_{2,n}\right)
	\Phi_{1,n}^{-1} Q_{n-1} 
- b  
 \left( I - \frac{1}{\zeta}\Phi_{1,n}^{-1} Q_{n-1}
 \Phi_{2,n-1}\right) \Phi_{1,n-1}^{-1} Q_{n-2} 
\nonumber \\ & 
+ b  \Phi_{1,n}^{-1} Q_{n-1} \left( \zeta \Phi_{2,n}
 -\frac{1}{\zeta}\Phi_{2,n-1} \right) \Phi_{1,n-1}^{-1} Q_{n-2} 
\nonumber \\
=& \;\, a \left(I-\frac{1}{\zeta} \Phi_{1,n+1}^{-1}Q_n \Phi_{2,n} \right)^{-1} 
 \left( \Phi_{1,n+1}^{-1} Q_n  - \zeta^2 \Phi_{1,n}^{-1} Q_{n-1} \right)
\nonumber \\ & 
+ b \left( I -\zeta \Phi_{1,n}^{-1} Q_{n-1} \Phi_{2,n}\right)
 \left( \frac{1}{\zeta^2} \Phi_{1,n}^{-1} Q_{n-1} 
 - \Phi_{1,n-1}^{-1} Q_{n-2} \right)
\end{align}
and 
\begin{align}
\left( \Phi_{2,n} \right)_t
&= 
\zeta a R_{n-1} \Phi_{1,n} + \frac{b}{\zeta} R_{n} \Phi_{1,n}
\left( I - \zeta\Phi_{1,n}^{-1} Q_{n-1} \Phi_{2,n}\right)
\nonumber \\
&= \zeta^2 a \left( \Phi_{2,n}-\frac{1}{\zeta^2}\Phi_{2,n-1} \right)
 \Phi_{1,n-1}^{-1} \Phi_{1,n} 
\nonumber \\ & \hphantom{=}\;
+ b \left( \Phi_{2,n+1}-\frac{1}{\zeta^2}\Phi_{2,n} \right)
\left( I- \zeta \Phi_{1,n}^{-1} Q_{n-1} \Phi_{2,n} \right)
\nonumber \\
&= a \left( \zeta^2 \Phi_{2,n}- \Phi_{2,n-1} \right)
\left(I- \frac{1}{\zeta}\Phi_{1,n}^{-1} Q_{n-1} \Phi_{2,n-1} \right)^{-1}
\nonumber \\ 
& \hphantom{=}\;
+ b \left( \Phi_{2,n+1}-\frac{1}{\zeta^2}\Phi_{2,n} \right)
\left( I- \zeta \Phi_{1,n}^{-1} Q_{n-1} \Phi_{2,n} \right), 
\label{AL-phi2}
\end{align}
\end{subequations}
respectively. 
Because \mbox{$\Phi_{1,n}^{-1} Q_{n-1}=q_n$} and \mbox{$ \Phi_{2,n}=r_n$}, 
(\ref{AL-phi}) can be identified with
(\ref{dCLL}). \hfill $\Box$


\section{\mbox{$(2+1)$}-dimensional Chen--Lee--Liu systems}
\label{app3}

In this appendix, we derive 
two 
\mbox{$(2+1)$}-dimensional 
generalizations of the Chen--Lee--Liu system 
from the Lax representations for
the two \mbox{$(2+1)$}-dimensional NLS systems 
considered 
in section~\ref{2D-PDE}. 
These results 
can be found by
exploiting the transformation of dependent variables
between
each pair of 
the Gerdjikov--Ivanov system
and the Chen--Lee--Liu system in \mbox{$2+1$} dimensions.

First, we discuss 
the non-isospectral 
case 
considered in subsection~\ref{sec-Zak}. 
Let $Q$ and $R$, together with
the auxiliary fields $f$ and $g$, 
satisfy the 
Calogero--Degasperis system (\ref{Z-sys}). 
Let $\Psi_1$ and $\Psi_2$ be the
first and second components of 
an eigenfunction
of the associated linear problem
(\ref{Za-UV}), wherein the $x$-independent spectral parameter 
$\zeta$ satisfies the non-isospectral condition 
\mbox{$\zeta_t = 2\zeta \zeta_y$}. 
Then, we have the following proposition.
\vspace{5mm}
\begin{proposition}
\label{propC.1}
{\em
The new pair of variables,
\begin{equation}
\left\{
\begin{split}
& q:= \left( \Psi_1 {\mathrm e}^{\mathrm{i} \zeta x} \right)^{-1} Q,
\\
& r:= \Psi_2 {\mathrm e}^{\mathrm{i} \zeta x},
\end{split}
\right.
\label{}
\end{equation}
satisfies 
the \mbox{$(2+1)$}-dimensional 
Chen--Lee--Liu system
%
\begin{equation}
\left\{
\begin{split}
& \mathrm{i} q_t + \mathrm{i} \zeta_y q 
+q_{xy} +2\mathrm{i}\zeta Aq + A q_x 
- \frac{1}{2} q \left[ B - (rq)_y\right] =O,
\\
& \mathrm{i} r_t + \mathrm{i} \zeta_y r
- r_{xy} -2\mathrm{i}\zeta r A 
+ r_x A + \frac{1}{2} \left[ B + (rq)_y\right] r =O,
\\
& A_x + \left[ qr,A \right]= (qr)_y,
\\
& B_x = \left( -4\mathrm{i}\zeta rq + r_x q - rq_x \right)_y,
\\
& \zeta_t = 2\zeta \zeta_y, \hspace{5mm} \zeta_x=0, 
\end{split}
\right.
\label{2DCLL}
\end{equation}
where $A$ and $B$ are new auxiliary 
fields. 
}
\end{proposition}
\vspace{5mm}
%
\noindent
{\bf Remark.}~~When $\zeta$ is a real function 
and the dependent variables are scalar, 
we can 
impose the complex conjugacy reduction 
\mbox{$r=\mathrm{i}\sigma q^\ast$}, 
\mbox{$A^\ast=-A$}, and \mbox{$B^\ast=B$} 
with a real constant $\sigma$. 
In particular, the reduction \mbox{$r=\mathrm{i}q^\ast$} 
simplifies (\ref{2DCLL}) to
\begin{equation}
\left\{
\begin{split}
& \mathrm{i} q_t + \mathrm{i} \zeta_y q 
+q_{xy} -2\zeta q \partial_x^{-1}\partial_y (|q|^2)
-2 q \partial_x^{-1}\partial_y (\zeta |q|^2)
\\
& \hspace{6mm}+\mathrm{i} q_x \partial_x^{-1}\partial_y (|q|^2)
- \frac{\mathrm{i}}{2} q \left[ \partial_x^{-1}\partial_y
  \left( q^\ast_x q - q^\ast q_x \right)
  - (|q|^2)_y\right] =0,
\\
& \zeta_t = 2\zeta \zeta_y, \hspace{5mm} \zeta_x=0.
\end{split}
\right.
\label{sCLL1}
\end{equation}
If $\zeta$ is a constant independent of $y$ and $t$, 
(\ref{sCLL1}) further reduces to
(cf.~(66) in~\cite{Myrz98} or (73) 
in~\cite{Myrz00})
\begin{align}
 \mathrm{i} q_t &+q_{xy} -4\zeta q \partial_x^{-1}\partial_y (|q|^2)
\nonumber \\ & 
+\mathrm{i} q_x \partial_x^{-1}\partial_y (|q|^2)
- \frac{\mathrm{i}}{2} q \left[ \partial_x^{-1}\partial_y
  \left( q^\ast_x q - q^\ast q_x \right)
  - (|q|^2)_y\right] =0.
\nonumber
\end{align}
%
Considering a gauge transformation 
of the Lax representation~\cite{Myrz00}, 
we can 
relate 
this \mbox{$(2+1)$}-dimensional Chen--Lee--Liu equation 
to the known \mbox{$(2+1)$}-dimensional Kaup--Newell equation 
[(4.3) in Ref.~\citen{Stra92}]. 
\vspace{5mm}
\\
\noindent
{\bf Proof.}~~For brevity, we
use the
quantities \mbox{$\Phi_j :=\Psi_j {\mathrm e}^{\mathrm{i} \zeta x} \; (j=1,2)$}
instead of $\Psi_j$. From (\ref{Za-U}), we 
obtain 
\mbox{$\Phi_{1,x}=Q\Phi_2$}
and \mbox{$R= \left(\Phi_{2,x}-2\mathrm{i}\zeta\Phi_2 \right) \Phi_1^{-1}$}. 
Thus, we can introduce the auxiliary field $A$ as 
\[
\Phi_{1,x} =  \Phi_1 \Phi_1^{-1} Q \Phi_2,\hspace{5mm}
\Phi_{1,y} =  \Phi_1 A. 
\]
The compatibility condition 
of 
these ``linear'' 
PDEs for $\Phi_1$ implies the relation 
\begin{equation}
\left( \Phi_1^{-1} Q \Phi_2 \right)_y -A_x
-\left[ \Phi_1^{-1} Q \Phi_2 , A \right] =O. 
\label{2D-CLL1}
\end{equation}
%
Equation (\ref{Z-g}) for the auxiliary field $g$ 
can be rewritten as 
\begin{equation}
g_x= \left( -2\mathrm{i}\zeta\Phi_2 \Phi_1^{-1}Q +\Phi_{2,x}\Phi_1^{-1}Q 
\right)_y.
\label{2D-CLL2}
\end{equation}
Using (\ref{Za1}) and (\ref{Za-V}), we can 
express 
the time derivatives of
\mbox{$\Phi_{1}^{-1}Q$} and $\Phi_{2}$ as
\begin{align}
\mathrm{i} \left( \Phi_1^{-1}Q \right)_t 
=& - \! \Phi_1^{-1}Q_{xy} + \Phi_1^{-1}Q g 
 -2\mathrm{i}\zeta \Phi_1^{-1}\Phi_{1,y}\Phi_1^{-1} Q 
 + \Phi_1^{-1}Q_y \Phi_{2}\Phi_1^{-1} Q 
\nonumber \\ 
=& - \left( \Phi_1^{-1}Q \right)_{xy} 
 - \left( \Phi_1^{-1} Q \Phi_2 \Phi_1^{-1} \right)_y Q 
 - \Phi_1^{-1} Q \Phi_2 \Phi_1^{-1} Q_y -A \Phi_1^{-1} Q_x  
\nonumber \\ &
+ \Phi_1^{-1}Q g 
 -2\mathrm{i}\zeta A \Phi_1^{-1} Q 
 + \left( \Phi_1^{-1}Q \right)_y \Phi_{2}\Phi_1^{-1} Q 
 + A \Phi_1^{-1}Q \Phi_{2}\Phi_1^{-1} Q 
\nonumber \\ 
=& - \left( \Phi_1^{-1}Q \right)_{xy} 
 - \Phi_1^{-1} Q \left( \Phi_2 \Phi_1^{-1} Q \right)_y 
 - A \left( \Phi_1^{-1} Q \right)_x  
 + \Phi_1^{-1}Q g 
\nonumber \\ &
 - \! 2\mathrm{i}\zeta A \Phi_1^{-1} Q
\label{2D-CLL3}
\end{align}
and 
\begin{align}
\mathrm{i} \Phi_{2,t}
&= 2 \mathrm{i}\zeta \Phi_{2,y} + R_y \Phi_1 -g \Phi_2
\nonumber \\
&= 2 \mathrm{i}\zeta \Phi_{2,y}  
+ \left[ \left( \Phi_{2,x}-2\mathrm{i}\zeta\Phi_2 \right) \Phi_1^{-1} 
	\right]_y \Phi_1 -g \Phi_2
\nonumber \\
&= \Phi_{2,xy }-2\mathrm{i}\zeta_y \Phi_2 
- \left( \Phi_{2,x}-2\mathrm{i}\zeta\Phi_2 \right) A -g \Phi_2,
\label{2D-CLL4}
\end{align}
%
respectively. 
By setting 
\mbox{$\Phi_1^{-1}Q=q$} and \mbox{$\Phi_2=r$}, 
we 
obtain 
a \mbox{$(2+1)$}-dimensional
system for $q$ and $r$, 
but it 
appears
asymmetric 
with respect to these variables. 
To 
rewrite it in a more symmetric form, 
we 
need only 
redefine 
the auxiliary field 
as 
\[
g =: -\mathrm{i}\zeta_y I + \frac{1}{2} (rq)_y + \frac{1}{2}B. 
\]
Thus, 
(\ref{2D-CLL1})--(\ref{2D-CLL4}), 
together with 
the non-isospectral condition, 
verify (\ref{2DCLL}).
\hfill $\Box$
\vspace{5mm}

Second, we discuss the 
case without 
a spectral parameter, 
considered in subsection~\ref{sec-DS}. 
Let $Q$ and $R$,  
together with 
the auxiliary fields $f$ and $g$, 
satisfy the Davey--Stewartson system (\ref{DS-sys}). 
Let $\Psi_1$ and $\Psi_2$ be the
first and second components
of a
solution
of the associated linear problem (\ref{DS-UV}). 
Then, we have the following proposition.
\vspace{5mm}
\begin{proposition}
\label{propC.2}
{\em
The new pair of variables,
\begin{equation}
\left\{
\begin{split}
& q:= \Psi_1^{-1} Q,
\\
& r:= \Psi_2,
\end{split}
\right.
\label{}
\end{equation}
satisfies
the \mbox{$(2+1)$}-dimensional 
Chen--Lee--Liu system
\begin{equation}
\left\{
\begin{split}
& \mathrm{i} q_t +a q_{xx} +bq_{yy} + q \left[ a(rq)_x +F\right] +2b Gq_y =O,
\\
& \mathrm{i} r_t -a r_{xx} -br_{yy} +\left[ a (rq)_x -F\right] r +2b r_y G=O,
\\
& F_y + a \left( r_y q - rq_y \right)_x =O,
\\
& G_x + \left[ qr, G\right]= (qr)_y,
\end{split}
\right.
\label{2DCLL2}
\end{equation}
where $F$ and $G$ are new auxiliary fields. 
}
\end{proposition}
\vspace{5mm}
\noindent
{\bf Remark.}~~In the case 
of scalar variables,
system (\ref{2DCLL2}) was studied in~\cite{SY97,MY97}. 
A special case of (\ref{2DCLL2}) for vector dependent variables 
was discussed in~\cite{Lou97}. 
\vspace{5mm}
\\
\noindent
{\bf Proof.}~~From (\ref{DS-U}), we have \mbox{$\Psi_{1,x}=Q\Psi_2$} 
and \mbox{$R= \Psi_{2,y}\Psi_1^{-1}$}. 
Thus, we can introduce the auxiliary field $G$ as
\[
\Psi_{1,x} =  \Psi_1 \Psi_1^{-1} Q \Psi_2,\hspace{5mm}
\Psi_{1,y} =  \Psi_1 G.
\]
The compatibility condition of 
these ``linear''
PDEs
for $\Psi_1$ implies the relation 
\begin{equation}
\left( \Psi_1^{-1} Q \Psi_2 \right)_y -G_x
-\left[ \Psi_1^{-1} Q \Psi_2 , G \right] =O.
\label{dDS1}
\end{equation}
Equation (\ref{DS-g}) for the auxiliary field $g$ 
can be 
rewritten as
\begin{equation}
g_y= 2a \left( \Psi_{2,y}\Psi_1^{-1}Q \right)_x.
\label{dDS2}
\end{equation}
Using (\ref{DS1}) and (\ref{DS-V}), we can
express
the time derivatives of
\mbox{$\Psi_{1}^{-1}Q$} and $\Psi_{2}$ as
\begin{align}
\mathrm{i} \left( \Psi_1^{-1}Q \right)_t
=& - \! a \Psi_1^{-1}Q_{xx} -b \Psi_1^{-1}Q_{yy} 
+\Psi_1^{-1}Qg 
+ b \Psi_1^{-1} \Psi_{1,yy} \Psi_1^{-1}Q 
\nonumber \\ & 
-a \Psi_1^{-1} Q \Psi_{2,x} \Psi_1^{-1} Q 
+a \Psi_1^{-1} Q_x\Psi_{2} \Psi_1^{-1} Q 
\nonumber \\[1mm]
=& - \! a \left( \Psi_1^{-1}Q \right)_{xx} 
- 2a \Psi_1^{-1}Q \Psi_2 \Psi_1^{-1}Q_{x} 
- a \left( \Psi_1^{-1}Q \Psi_2 \right)_x \Psi_1^{-1}Q 
\nonumber \\ & 
+ a \left( \Psi_1^{-1}Q \Psi_2 \right)^2 \Psi_1^{-1}Q 
- b \left( \Psi_1^{-1}Q \right)_{yy} - 2 b G \Psi_1^{-1}Q_y
\nonumber \\ & 
+ b G^2 \Psi_1^{-1}Q + \Psi_1^{-1}Q g 
+ b G^2 \Psi_1^{-1} Q 
- a \Psi_1^{-1} Q \Psi_{2,x} \Psi_1^{-1} Q 
\nonumber \\ & 
+ a \left( \Psi_1^{-1} Q \right)_x \Psi_2 \Psi_1^{-1} Q 
+ a \Psi_1^{-1} Q \left( \Psi_2 \Psi_1^{-1} Q \right)^2 
\nonumber \\[1mm]
=& - \! a \left( \Psi_1^{-1}Q \right)_{xx} 
- b \left( \Psi_1^{-1}Q \right)_{yy} 
- 2a \Psi_1^{-1}Q \Psi_2 \left( \Psi_1^{-1}Q \right)_{x} 
\nonumber \\ & 
- 2a \Psi_1^{-1}Q \Psi_{2,x} \Psi_1^{-1}Q + \Psi_1^{-1}Q g 
- 2 b G \left( \Psi_1^{-1}Q \right)_y
\label{dDS3}
\end{align}
and
\begin{align}
\mathrm{i} \Psi_{2,t}
&=  a \Psi_{2,xx} -g \Psi_2 -b R \Psi_{1,y} + b R_y\Psi_1 
\nonumber \\
&= a \Psi_{2,xx}  + b \Psi_{2,yy} -g \Psi_2 -2b R \Psi_{1,y} 
\nonumber \\
&= a \Psi_{2,xx}  + b \Psi_{2,yy} -g \Psi_2 -2b \Psi_{2,y} G, 
\label{dDS4}
\end{align}
respectively.
By setting
\mbox{$\Psi_1^{-1}Q=q$} and \mbox{$\Psi_2=r$},
we
obtain
a \mbox{$(2+1)$}-dimensional system for $q$ and $r$,
but it
appears
asymmetric
with respect to these variables.
To
rewrite it in a more symmetric form,
we 
redefine
the auxiliary field 
as
\[
g =: a (rq)_x -F.
\]
Thus,
(\ref{dDS1})--(\ref{dDS4}) verify (\ref{2DCLL2}).
\hfill $\Box$

\end{document}